\documentclass[aps,prb,superscriptaddress,twocolumn,10pt]{revtex4-2}

\usepackage{graphicx}
\usepackage{dcolumn}
\usepackage{bm}

\usepackage{amsmath}

\usepackage{graphicx}

\usepackage{relsize} 

\DeclareFontFamily{U}{mysize}{} 
\DeclareFontShape{U}{mysize}{m}{n}{ <-> s*[0.5] cmr10 }{} 

\usepackage{adjustbox}

\usepackage{xcolor}

\usepackage{caption}

\usepackage{comment}

\begin{document}

\preprint{APS/123-QED}

\title{Tracking the evolution of self-crossing points in memristive hysteresis loops}


\date{\today}

\title{Phase-Topology Classification of Memristor Hysteresis Loops via Self-Crossings}

\author{Ovidiu-Zeno Lipan}
\affiliation{Department of Physics, University of Richmond, 28 Westhampton Way, Richmond, Virginia 23173, USA}
\email{olipan@richmond.edu}

\author{Eric Neuhaus}
\affiliation{Department of Physics, University of Richmond, 28 Westhampton Way, Richmond, Virginia 23173, USA}

\author{Rafael Schio Wengenroth Silva}
\affiliation{Departamento de Física, Universidade Federal de São Carlos, 13565-905 São Carlos, SP, Brazil}

\author{Soumen Pradhan}
\affiliation{Julius-Maximilians-Universität Würzburg, Physikalisches Institut and Würzburg-Dresden Cluster of Excellence ct.qmat, Lehrstuhl für Technische Physik, Am Hubland, 97074 Würzburg, Deutschland}

\author{Fabian Hartmann}
\affiliation{Julius-Maximilians-Universität Würzburg, Physikalisches Institut and Würzburg-Dresden Cluster of Excellence ct.qmat, Lehrstuhl für Technische Physik, Am Hubland, 97074 Würzburg, Deutschland}

\author{Leonardo K. Castelano}
\affiliation{Departamento de Física, Universidade Federal de São Carlos, 13565-905 São Carlos, SP, Brazil}

\author{Ana Luiza Costa Silva}
\affiliation{Departamento de Física, Universidade Federal de São Carlos, 13565-905 São Carlos, SP, Brazil}

\author{Sven Höfling}
\affiliation{Julius-Maximilians-Universität Würzburg, Physikalisches Institut and Würzburg-Dresden Cluster of Excellence ct.qmat, Lehrstuhl für Technische Physik, Am Hubland, 97074 Würzburg, Deutschland}

\author{Victor Lopez-Richard}
\affiliation{Departamento de Física, Universidade Federal de São Carlos, 13565-905 São Carlos, SP, Brazil}

\date{\today}

\begin{abstract}
Memristive devices have revolutionized non-volatile memory and neuromorphic computing, yet the geometry of their hysteresis loops --- in particular, the occurrence and robustness of multiple self-crossings --- remains poorly understood. Here we introduce a topological and algebraic framework that treats the number of transverse self-intersections of a memristor hysteresis loop as a robust integer-valued invariant. Drawing on differential topology, singularity theory, and cusp catastrophe, we employ discriminants and resultants to stratify the six-dimensional parameter space. This approach partitions the parameter space into structurally stable regions separated by explicitly computable catastrophe surfaces. We demonstrate that the crossing number remains strictly invariant under continuous deformations and changes only at self-tangencies or cusp singularities, thereby providing a complete classification of all multi-lobed hysteresis behaviors. These insights bridge device physics with modern singularity theory and suggest a clear roadmap for exploiting higher-order memory effects in next-generation electronics and brain-inspired hardware.
\end{abstract}

\maketitle

\section{Introduction}

Although practical memory devices~\cite{Strukov2008} lack the magnetic induction strictly required by ideal memristors~\cite{Vongehr2015}, they successfully satisfy the broader criteria of generalized memristive systems~\cite{Chua1976}. Modern solid-state analysis further allows broadening this perspective toward macroscopic relaxation dynamics, where the apparent complexity of diverse microscopic mechanisms is captured by fundamental kinetic properties such as charge trapping rates and relaxation times~\cite{lopez2023tuning}. In this framework, the physical origin of the hysteresis becomes secondary to the topological properties of the current--voltage loop itself --- and, in particular, to the number and arrangement of its self-crossings.

Experimental evidence from oxide-based devices firmly establishes that these topological features are highly sensitive to control parameters and can undergo drastic transitions. Kubicek et al.\ demonstrated that varying the sweep voltage amplitude and rate in epitaxial SrTiO$_{3-\delta}$ thin films tunes the crossing count between one and three, driven by two competing switching mechanisms of opposite polarity~\cite{Kubicek2015}. Messerschmitt et al.\ showed that atmospheric moisture can transform the I--V profile of Pt$/$SrTiO$_3/$Pt devices from a pinched hysteresis into a capacitive trace without crossings, or induce complex trajectories with multiple crossings~\cite{Messerschmitt2015}. Analogous topological transitions have been reported in planar ZnO:Na memristors, where both the voltage sweep period and the gaseous environment systematically reshape the loop geometry~\cite{CostaSilva2025}. The diversity of microscopic origins underlying these observations --- bulk ionic, interfacial, and surface-driven REDOX --- renders device-specific kinetic models impractical for capturing the full phenomenology \cite{Li2014,Sah2015,Sun2020}. A generalized, mechanism-independent mathematical framework capable of classifying crossing-number transitions through computable bifurcation boundaries in control-parameter space is therefore essential.

Recent analyses utilizing impedance spectroscopy suggest that nonzero self-crossings in memristive I--V curves arise from transitions in the character of the hysteresis --- from capacitive-like to inductive-like --- driven by the delayed adaptation of the internal state variable~\cite{Bisquert2024c}. This phenomenon is not restricted to a single material platform; a unified phenomenological framework has shown that the same interplay between fast capacitive charging and slow state-variable dynamics underlies the hysteresis observed in halide perovskite solar cells, oxide memristors, organic electrochemical transistors, and biological ion channels~\cite{Bisquert2024b}. Bisquert et al.\ conjectured that such crossing behavior constitutes a general property of systems with self-crossing hysteresis loops~\cite{Bisquert2024c}; however, a rigorous mathematical classification determining how many crossings are possible and predicting the parametric boundaries where their number changes remains to be established.

In the present work, we develop such a classification. Starting from a generalized memristive system whose nonlinear response is captured by a polynomial generating function, we show that the number of self-crossings is determined by the roots of a voltage-dependent susceptance polynomial, and that the control-parameter space is partitioned into well-defined regions of constant crossing number separated by explicitly computable bifurcation boundaries. The underlying mathematical structure is that of a cusp catastrophe, which provides the rigorous foundation for this partitioning.

The resulting framework is directly applicable without requiring familiarity with catastrophe theory itself. The analysis yields explicit phase diagrams and a parameter-independent consistency test that, given only measured current--voltage data, determine how many crossings a device will exhibit and precisely where in parameter space their number changes---regardless of whether the underlying memory mechanism is bulk ionic, interfacial, or of any other microscopic origin.

\section{Generalized Memristive Systems: Capturing Non-Ideal Device Behavior}

The theoretical description adopted in this work is grounded in the framework
of generalized memristive systems introduced by Chua and
Kang~\cite{Chua1976}, rather than in the original ideal-memristor postulate
based on a constitutive relation between charge and magnetic flux linkage. In
the generalized formulation, the instantaneous current through a
voltage-controlled device takes the form
\begin{equation}
I(t) = W(n,V)\,V(t), \quad \dot{n} = f(n,V),
\label{eq:generalized}
\end{equation}
where $W$ is a state- and voltage-dependent memductance and $n$ denotes an
internal state variable whose dynamics are governed by the evolution
equation~$f$. A physically transparent realization of this
structure emerges from the microscopic theory of nonequilibrium carrier
transport in solids~\cite{Silva2022,lopez2023tuning}: the internal state
corresponds to the density of charge carriers trapped at nonequilibrium sites,
and memory arises from the finite inertia of the trapping and detrapping
processes, which prevents the internal state from following the applied
stimulus instantaneously. A characteristic relaxation time~$\tau$ governs how
quickly this adjustment proceeds---the larger~$\tau$, the more persistent the
memory.

The behavior of a memristive device under periodic excitation is governed by
the interplay between two complementary aspects of the external drive: the
instantaneous voltage~$V$, which dictates the electronic configuration that the
system would adopt at equilibrium, and its time derivative~$dV/dt$, which
measures how rapidly that configuration is being swept. While $V$ sets the
target toward which the internal state tends, $dV/dt$ quantifies the rate at
which this target shifts---ensuring that, for finite relaxation time~$\tau$, the
internal state perpetually lags behind the instantaneous drive. The relative
weight of these two ingredients is captured by the dimensionless parameter
$p = \omega\tau$, the product of the driving angular frequency and the
relaxation time. Here $\tau$ plays the role of an inertia: the larger its value,
the more slowly the internal state responds to any new directive---whether
a freshly imposed target from the drive or the erasure of a previous
one---so that $p$ effectively measures the competition between the rate at which
new commands arrive and the rate at which the system can process
them. When $p \ll 1$ the system tracks the voltage adiabatically and
no hysteresis appears; when $p \gg 1$ the internal state is effectively frozen
and the device responds as a linear resistor. Memory effects---and with them
the self-crossings of the current--voltage loop---are most prominent in the
intermediate regime where $p \sim 1$, a condition previously identified as
optimal for the memristive response~\cite{Silva2022}.

Together, the pair $(V,\,dV/dt)$ defines a natural phase space for the external
stimulus: the first coordinate specifies the instantaneous target imposed on
the internal state, while the second measures how rapidly that target is
displaced. For a sinusoidal drive $V(t) = A\cos(\omega t)$, the trajectory in
this space is a closed curve---an ellipse that, after appropriate
normalization, maps onto the unit circle. In this representation, one axis is
proportional to the applied voltage and the other to its rate of change, so
that each point on the circle encodes both \textit{where} the drive directs the
system and \textit{how urgently} it is being redirected. The current measured at
each such point depends not only on the present voltage but also on the
accumulated history encoded in the internal state. This geometric viewpoint
provides a clean separation of the ascending and descending branches and sets
the stage for the algebraic analysis of self-crossing points developed in the
following sections.

It is worth noting that the pair $(V,\,dV/dt)$ adopted here is not the only
natural parametrization of the stimulus cycle. Under sinusoidal excitation,
$dV/dt$ and the time-integral $\int V\,dt$ are strictly proportional, so that
either coordinate maps onto the same unit-circle topology after normalization.
Since $\int V\,dt$ coincides with Chua's original flux
variable~\cite{Chua1971} --- stripped of its magnetic connotation, as argued by
Vongehr and Meng~\cite{Vongehr2015} --- the topological classification of
self-crossings developed in this work is independent of which conjugate
variable is chosen to complement the voltage.

To enable a systematic search for hysteresis loops exhibiting multiple
self-crossing points, the present work adopts a specific, analytically
tractable subclass of generalized memristive systems~\eqref{eq:generalized}
by reducing the generality of the state-evolution and memductance functions
while preserving the essential voltage-controlled physics.

\section{\label{sec:THETA_Zero}Voltage-Controlled Memristive Model: Relaxation Time, Conserved Quantities, and Parameter-Independent Validation}

A concise yet powerful formulation for $f(n, V)$ that retains the ability to capture state evolution driven by voltage in a wide range of experimental settings \cite{lopez2023tuning} is obtained by explicitly introducing a relaxation time $\tau$
\begin{align}\label{f_g}
f(n, V) = -\frac{n}{\tau} + g(V),
\end{align}
where $g(V)$ remains unspecified for the time being. As for the memductance, the approach is to regard it as a perturbed constant admittance $G$, modulated by the state variable $n$, such that
\begin{align}\label{W_G}
W = G + n.
\end{align}
This reduced formulation strikes an effective balance between analytical tractability and physical relevance. Before investigating the conditions under which it produces hysteresis loops with multiple self-crossing points---the central objective of this work---it is essential to establish whether the model can faithfully reproduce experimentally observed memristive behavior.
Given a series of measured $(I, V)$ data pairs that exhibit a hysteresis loop, we evaluate how well the model defined by Eqs.~(\ref{f_g}) and (\ref{W_G}) captures the observed dynamics.
To this end, we derive a general relationship satisfied by the $(I, V)$ points that is independent of the specific functional form of $g(V)$.
This procedure constitutes an essential parameter-independent consistency test. It provides a means to assess whether the reduced generalized model adequately describes the observed hysteresis behavior of a given physical device. At the same time, the relation elucidates the distinct dynamics governing the ascending and descending branches of the hysteresis loop, thereby providing the foundation for the analysis that follows.
Applying a voltage waveform that increases from $-A$ to $A$ (ascending or positive branch) and subsequently decreases from $A$ to $-A$ (descending or negative branch), with $A$ denoting the amplitude, traces a hysteresis loop in the $I$--$V$ plane consisting of two distinct branches, labeled $+$ and $-$, respectively.
The dynamics on each branch are governed by
\begin{align}
\frac{d n_+}{dt} + \frac{n_+}{\tau} &= g(V), \qquad
\frac{dV}{dt} = h_+(V), \label{eq:model1} \\
\frac{d n_-}{dt} + \frac{n_-}{\tau} &= g(V), \qquad
\frac{dV}{dt} = h_-(V), \label{eq:model3}
\end{align}
together with the current--voltage relations
\begin{align}
I_+ &= (G + n_+) V, \qquad
I_- = (G + n_-) V. \label{eq:IVconnection}
\end{align}
Because the input voltage is externally controlled, the sweep rates $h_{\pm}(V) \equiv dV/dt$ are known functions. It is therefore convenient to change the independent variable from time $t$ to voltage $V$, yielding the state equations
\begin{align}
h_+(V)\frac{d n_+}{dV} + \frac{n_+}{\tau} &= g(V), \\
h_-(V)\frac{d n_-}{dV} + \frac{n_-}{\tau} &= g(V).
\end{align}
Since both branches share the same single-valued generating function $g(V)$, we immediately obtain the relation
\begin{align}\label{eq:conserved1}
h_+(V)\frac{d n_+}{dV} + \frac{n_+}{\tau}
= h_-(V)\frac{d n_-}{dV} + \frac{n_-}{\tau},
\end{align}
which holds independently of $g(V)$ and can be regarded as a conserved quantity along the hysteresis loop.
Expressing the state variable in terms of measurable quantities,
\begin{align}
n_{\pm}(V) = \frac{I_{\pm}(V)}{V} - G,
\end{align}
and substituting into (\ref{eq:conserved1}) yields a general relationship involving only experimental data $(I,V)$:
\begin{align}\label{eq:conserved2}
\tau \, h_+(V) \frac{d}{dV}\left(\frac{I_+}{V}\right) + \frac{I_+}{V}
= \tau \, h_-(V) \frac{d}{dV}\left(\frac{I_-}{V}\right) + \frac{I_-}{V}.
\end{align}
(Note that this relation is singular at $V=0$, consistent with the pinched nature of the loop at the origin.) 

This is precisely the $g(V)$-independent test promised earlier and serves as a powerful consistency check: any experimental hysteresis loop that satisfies (or closely approximates) (\ref{eq:conserved2}) for a time constant $\tau$ that is either independently estimated (via step-response) or extracted statistically from the data (appearing relatively constant across the loop) is compatible with the reduced model.

Once the model has been validated with the consistency test, the generating function can be reconstructed directly from either the ascending or the descending branch. For the ascending branch it reads
\begin{align}\label{eq:gvIV}
g(V) = h_+(V) \frac{d}{dV}\left(\frac{I_+}{V}\right) + \frac{1}{\tau}\left(\frac{I_+}{V} - G\right).
\end{align}

In what follows, we assume that, for the device under consideration, the generating function $g(V)$ can be approximated by a fourth-order polynomial. The constant (zero-order) term of this polynomial is absorbed into the baseline admittance $G$, so that $g(V)$ begins with a linear term in voltage. We then adopt the scaled dimensionless form
\begin{align}\label{ScaledGeneratingFunction}
    \frac{\tau}{G} g(V) = c_1 V + c_2 V^2 + c_3 V^3 + c_4 V^4
\end{align}
to facilitate the subsequent perturbation analysis with respect to $G$ and the time constant $\tau$.

\section{Canonical Unit-Circle Framework for Multi-Crossing Analysis of Sinusoidally Driven Memristive Hysteresis }

The device is driven by a cosine voltage of amplitude $A$ and dimensionless angular frequency $p = \omega \tau$. Introducing the dimensionless time variable $t' = t / \tau$ yields
\[
\cos(\omega t) = \cos(p t').
\]
For notational convenience, we drop the prime and write the applied voltage as
\[
V(t) = A \cos(p t),
\]
where $t$ now denotes dimensionless time.

\begin{figure}[h]
\includegraphics[width=0.48\textwidth]{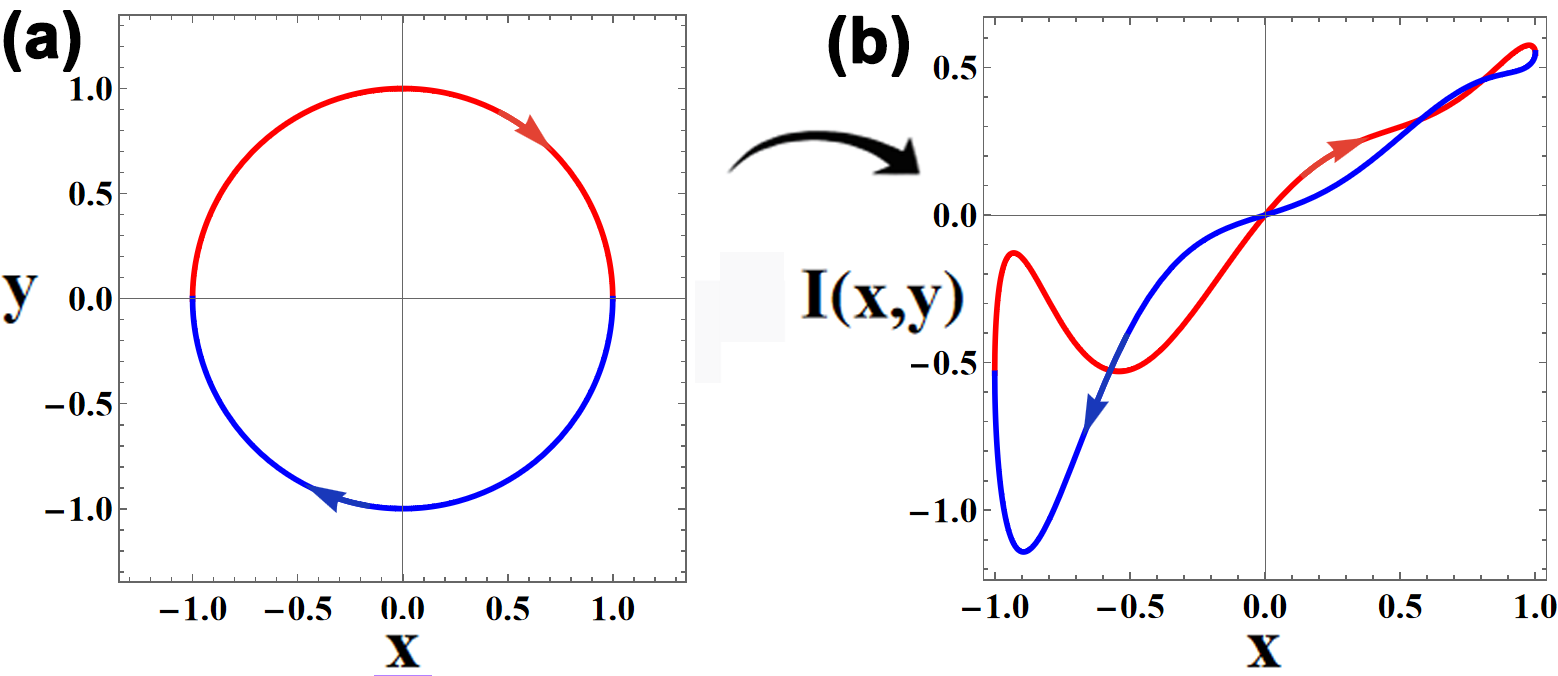}
\caption{\label{fig:LoopCircle} Mapping of the current–voltage hysteresis loop from the unit circle.
(a) Unit circle in the normalized voltage--derivative plane. The horizontal axis shows the normalized voltage \(x = V/A\), while the vertical axis shows its time derivative \(y\). As time progresses over one full period, the point \((x(t), y(t))\) traces the entire circle exactly once.
(b) Resulting current--voltage hysteresis loop \(I(x,y)\). Each point on the circle in (a) is mapped to a point on the loop in (b) at the same instant in time. The upper semicircle (red) maps to the ascending (positive) branch of the hysteresis loop, while the lower semicircle (blue) maps to the descending (negative) branch. Arrows indicate the direction of traversal with time. The hysteresis is  generated using $\tau=1$ and, $c_1=-15.7$, $c_2=35.7$, $c_3=33.0$, $c_4=-66.7$   where $A=0.77$ and $p=0.879$}
\end{figure}

The resulting hysteresis loop appears as a closed parametric curve in the $I$--$V$ plane, with time serving as the parameter. In this trigonometric setting, the clock-wise oriented unit circle $(x,y)=(\cos(p t), -\sin(p t))$ is the canonical example of a closed parametric curve. It is therefore natural to represent the hysteresis loop as the image of the unit circle under a suitable mapping, so that traversal along the circle corresponds directly to traversal along the hysteresis loop. This mapping is illustrated in Fig.~\ref{fig:LoopCircle}.

Although the $I$--$V$ representation is intuitive and widely used, the introduction of the unit circle in an auxiliary $(x,y)$-plane provides a particularly useful geometric framework for cleanly separating the ascending and descending branches.

To enable a direct analysis of multiple self-crossing points---the central objective of this work---we embed the loop in this plane via the following canonical mapping, in which the normalized voltage is defined as $x = V/A = \cos(p t)$
\[
(\cos(p t), -\sin(p t)) \mapsto (x, y).
\]

The auxiliary coordinate then satisfies $y = \pm \sqrt{1 - x^2}$, where the choice of sign distinguishes the two branches: $y < 0$ for the descending sweep (voltage decreasing from $+A$ to $-A$) and $y > 0$ for the ascending sweep (voltage increasing from $-A$ to $+A$).

In this construction the current becomes a function $I(x,y)$ defined on the $(x,y)$-plane. Evaluating $I$ along either branch $y = \pm \sqrt{1 - x^2}$ immediately recovers the corresponding branch of the experimental hysteresis loop. 

In addition, treating $x$ and $y$ as independent variables rather than restricting them to the circle $y = \pm \sqrt{1 - x^{2}}$ provides a second, complementary viewpoint on $I(x,y)$. This dual perspective proves useful in defining the admittance and related response functions of the system. The construction is reminiscent of the step from real numbers to the complex plane $(1,i)$ in circuit theory; here, we instead move  to the two-dimensional $(x,y)$-plane, where the loop appears as a curve that can be studied geometrically.

With the $(x,y)$-plane representation now established, we solve the reduced generalized model for the cosine driving voltage $V(t) = A \cos(p t)$. Chebyshev polynomials \cite{Mason2003} of the first and second kind are particularly well suited to this nonlinear problem: they express every higher harmonic $\cos(m p t)$, where $m$ is an integer, as a polynomial in $x = \cos(p t)$ alone, while $\sin(m p t)$ appears linearly in $y$ multiplied by a polynomial in $x$.

The explicit solution of the reduced model under the cosine drive is a fourth-order polynomial in the normalized variables $x$. The full symbolic expression is too long to show in its entirety on a single page, but it is straightforward to manipulate using Mathematica \cite{Mathematica}.  Only the contributions up to quadratic order in the drive amplitude $A$ (i.e., those involving $c_1$ and $c_2$) are displayed below; the remaining higher-order terms (involving $c_3$ and $c_4$) follow the identical structure and are required for the full description of the loop:
\begin{align}\label{eq:IVscaled}
\begin{split}
\frac{I}{A G}=x \left( 1+ A c_1 \frac{x}{1 + p^2} + A^2 c_2 \frac{2 p^2 + x^2}{1 + 4 p^2}+\cdots \right) \\
+ x \left( -A c_1 \frac{p}{1 + p^2} - A^2 c_2 \frac{2 p x}{1 + 4 p^2}+\cdots \right) y
\end{split}
\end{align}
The scaled generating function \eqref{ScaledGeneratingFunction} proves particularly convenient here: it normalizes the constant admittance to unity, so that both current and voltage appear in dimensionless form ($I/(A G)$ versus $x = V/A$). The free parameters separate naturally into two extrinsic quantities---the drive amplitude $A$ and the dimensionless frequency $p$---and the four intrinsic coefficients $c_k$ ($k=1,\dots,4$).

This compact polynomial expression constitutes the stepping stone for the subsequent analysis. It supplies the explicit algebraic structure needed to derive the precise conditions under which the hysteresis loop exhibits multiple self-crossing points---the central objective of this work.

\section{Susceptance Polynomial Roots as Origin of Multiple Hysteresis Crossings}

 By assigning the current according to the sign of the auxiliary variable $y$, both branches of the hysteresis loop can be unified into the single compact expression
\begin{align}\label{eq:Ixy}
\frac{I(x,y)}{AG} = P_X(x) + P_Y(x)\, y,
\end{align}
where $P_X(x)$ and $P_Y(x)$ are polynomials that depend exclusively on $x$.

The crossing points, where the ascending and descending branches intersect ($I_+=I_-$), satisfy
\begin{align}
P_X(x) + P_Y(x)\, y = P_X(x) + P_Y(x)\, (-y).
\end{align}

All self-crossing points occur at the roots of the polynomial equation
\begin{align}
P_Y(x) = 0.
\end{align}

We examine the detailed structure and multiplicity of these crossings in the following. However, it is important to note that the present physical model, being formulated for a passive device, is valid only within restricted ranges of the drive amplitude $A$ and dimensionless frequency $p$ such that the entire hysteresis loop remains confined to the first and third quadrants of the $I$--$V$ plane.

In terms of voltage-dependent conductance $G(x)$, defined by $P_X(x)=G(x) x$, and the susceptance $B(x)$, defined by $P_Y(x)=B(x) x$, this confinement condition can be expressed as follows:
\begin{align}\label{PassiveDevice}
    -G(x)\leq B(x) \sqrt{1-x^2} \leq G(x)
\end{align}

\begin{figure}[h]
    \centering
    \begin{tabular}{c@{\hspace{1mm}}c} 
        \includegraphics[width=0.23\textwidth]{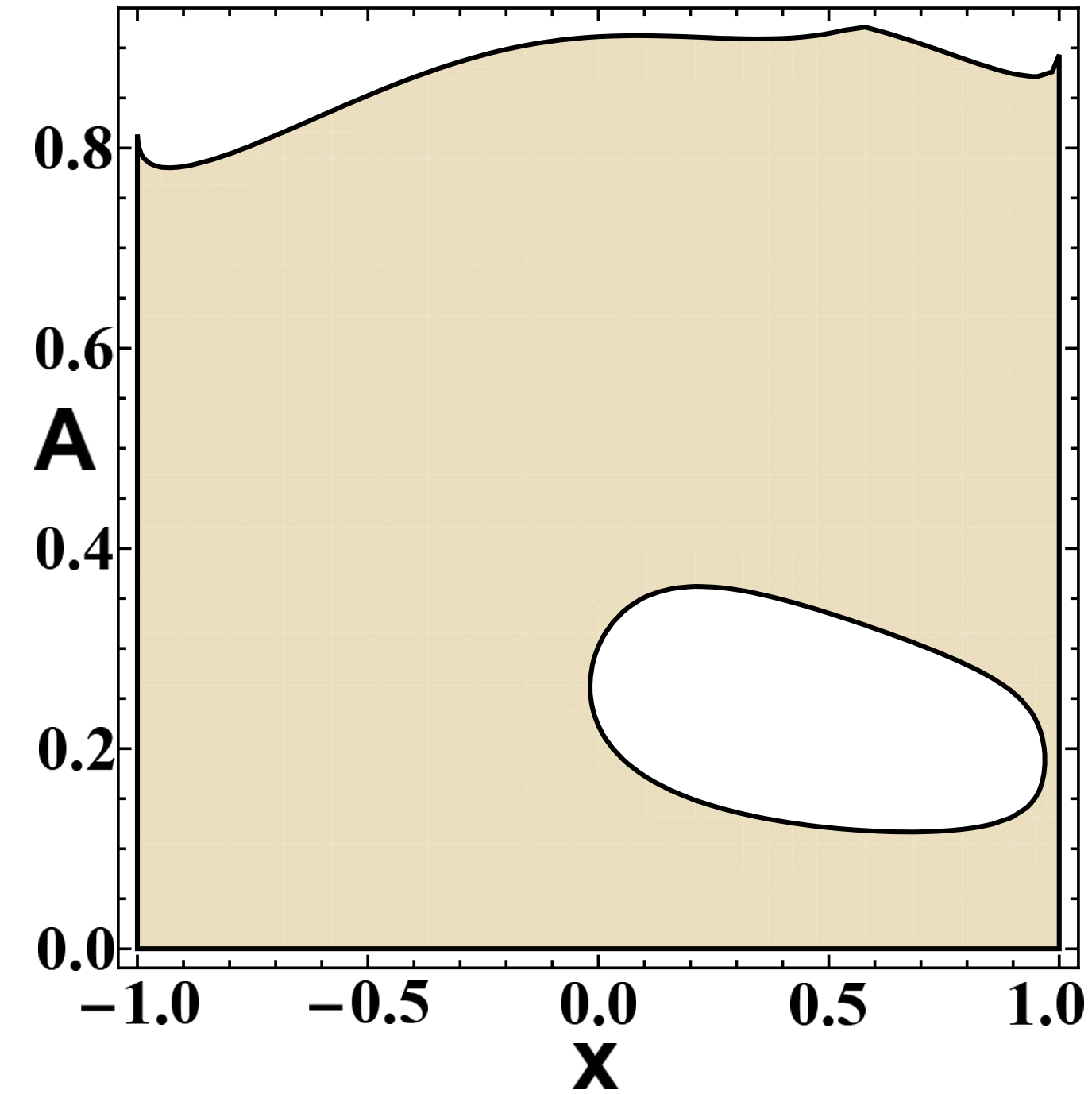} &
        \includegraphics[width=0.235\textwidth]{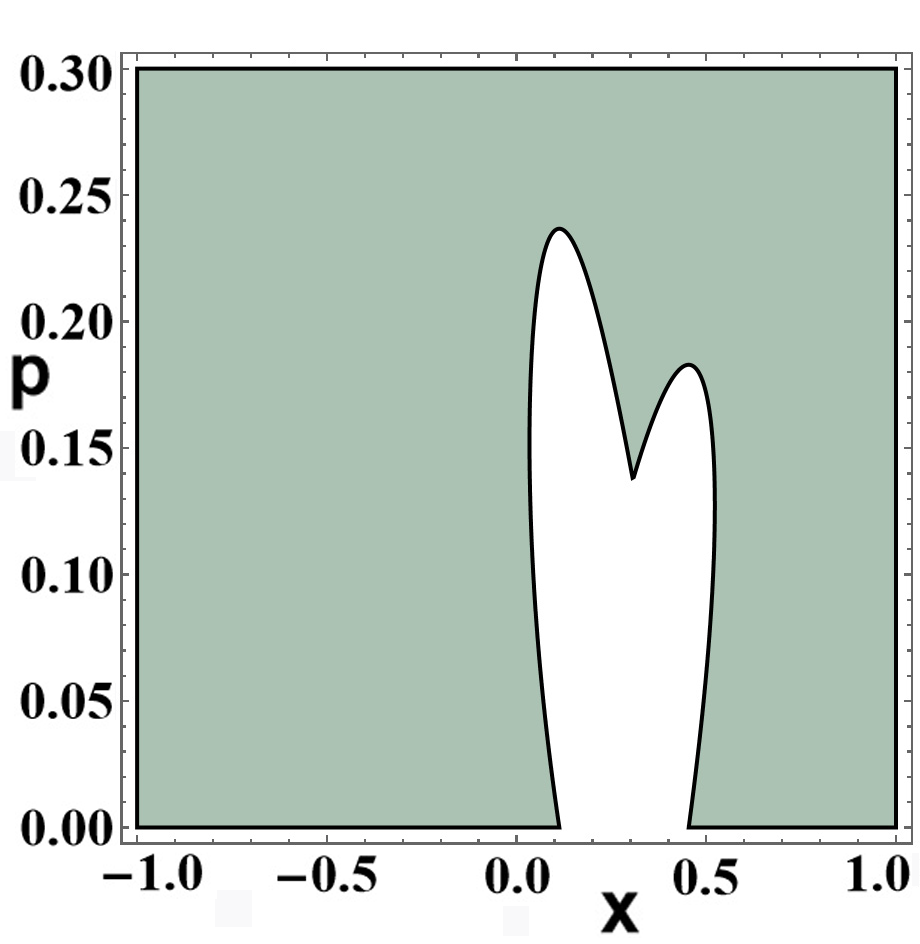}\\
    (a) $p=0.879$  &
    (b) $A=0.70$ 
    \end{tabular}
    \caption{\label{fig:CombinedFigures}Constraints for hysteresis loops confined to quadrants I and III in the I-V plane. The passivity constraint (\ref{PassiveDevice}) is satisfied in the colored regions and violated in the white regions. Both panels are generated using $\tau=1$ and the same generating function as in  Fig.~\ref{fig:LoopCircle}.}
\end{figure}
Fig.\ref{fig:CombinedFigures} (a) underscores the importance of carefully assessing the generating function model, as there is no assurance that a model valid for one amplitude will work for all smaller values. For example, an amplitude of 0.2 proves unsuitable, as the loop extends into quadrant IV for positive input voltage, violating the condition of confinement to quadrant I. This issue persists within a range of amplitudes near 0.2. Furthermore, amplitudes above 0.79 result in the loop extending into quadrant II for large negative voltages, breaking the confinement to quadrant III. Fig.\ref{fig:CombinedFigures} (b) demonstrates a similar effect when varying frequency, indicating that the model is ineffective for very low frequencies at a fixed amplitude.

This conclusion highlights a rather straightforward fact: nonlinear models are only valid within a specific range of amplitude and frequency, and cannot be extended beyond that region. Yet, this constraint is not entirely detrimental; it also permits the coefficients $c_k$
to be significantly larger than one, as in Fig.\ref{fig:CombinedFigures}, since the influence of higher-order terms diminishes as powers of a small amplitude reduce their weight. 

\begin{figure}[h]
\includegraphics[width=0.35\textwidth]{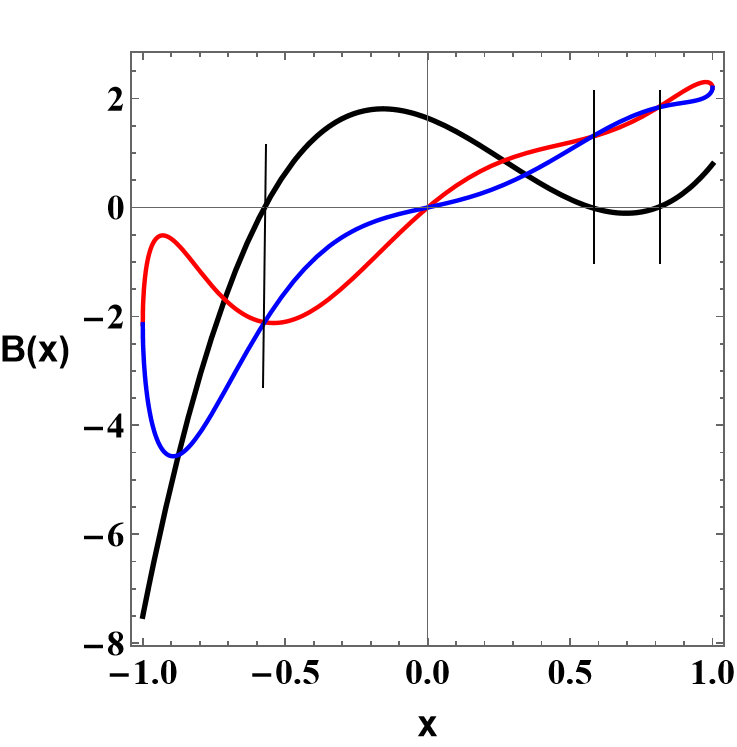} \caption{\label{fig:HysteresisIVloop}Relationship between the susceptance \( B(x) \) (black) and the self-crossings of the current--voltage hysteresis loop (red and blue, same as in Fig.~\ref{fig:LoopCircle}). The thin vertical segments indicate the roots of \( B(x) \), which mark the voltages at which the hysteresis loop crosses itself (besides the crossing at zero-voltage). }
\end{figure}

Returning to the crossings, the choice of the generating function coefficients $c_k$
in Fig.\ref{fig:CombinedFigures} is also motivated by the existence of admissible amplitude–frequency pairs for which the hysteresis loop exhibits three crossings, as shown in Fig.\ref{fig:HysteresisIVloop}.
The red curve in Fig.\ref{fig:HysteresisIVloop} corresponds to an increasing voltage sweep from $-A$ to $A$, while the blue curve corresponds to the decreasing sweep.

The susceptance $B(x)$, shown in black in Fig.~\ref{fig:HysteresisIVloop}, is a third-order polynomial that crosses the horizontal axis three times within the voltage range $-A < V < A$. Each intersection corresponds to a self-crossing of the hysteresis loop. Dividing by 
$x$ in 
$B(x)=P_Y(x)/x$ removes the inherent crossing at the origin, which is present by default in \eqref{eq:generalized}.

This result places on rigorous algebraic footing the conjecture of Bisquert et al.~\cite{Bisquert2024c} that self-crossings are a general property of memristive hysteresis loops, and further establishes that their number is bounded by the degree of $B(x)$.

Beyond its critical role in locating multiple self-crossing points, the nonlinear voltage-dependent susceptance \(B(x)\) offers deeper physical insight into the reactive dynamics of the memristor and opens additional methods for device characterization and design.

One benefit of expressing the susceptance $B(x)$ as a polynomial in the voltage $x = V/A$ lies in the direct evaluation of the local admittances, defined as the slopes of the $I$-$V$ curve. This approach simplifies the analysis of the admittance’s distinction between the two branches of the hysteresis loop as the voltage varies, both at arbitrary voltages and specifically at the crossings. Specifically, with the current written as 
\begin{align}
    I_{\pm}= x G(x)+x B(x) y_{\pm}(x)
\end{align}
where $y_{\pm}(x)=\pm \sqrt{1-x^2}$ we get
\begin{align}
\left. \frac{dI_{+}}{dx} - \frac{dI_{-}}{dx} \right|_{x=0}=2 B(0).
\label{eq:B0}
\end{align}
The magnitude of $B(0)$ reflects the extent of the difference between the on- and off-state admittances at zero voltage, while its sign determines the orientation of the loop. When $B(0) = 0$, the hysteresis is classified as type~II.
Expressed in terms of the generating-function coefficients, this becomes
\begin{align}
2 B(0)=-2 p \left( \frac{A}{(1 + p^2) }c_1 + \frac{6 A^3 p^3}{(1 + p^2)(1 + 9 p^2) } c_3\right).
\end{align}
Interestingly, at a crossing point with  $B(x_c) = 0$ and nonzero voltage $x_c$, we can still establish a simple relation

\begin{align}\label{SlopesAtCrossing}
\left. \frac{dI_{+}}{dx} - \frac{dI_{-}}{dx} \right|_{x_c}=2 x_c \sqrt{1 - x_c^2} \left. \frac{dB}{dx} \right|_{x=x_c}
\end{align}

When reformulated in terms of the generating-function coefficients, the left-hand side of (\ref{SlopesAtCrossing}) for this case becomes:
\begin{align}
-2 p x_c \sqrt{1 - x_c^2} \left( \frac{2 A^2}{1 + 4 p^2}c_2 + \frac{6 A^3 x_c}{1 + 9 p^2}c_3 \right. & \notag \\
\left. + \frac{12 A^4 \left( 2 p^2 + (1 + 4 p^2) x_c^2 \right)}{(1 + 4 p^2)(1 + 16 p^2)}c_4 \right)
\end{align}
This term may vanish, implying that the two branches are tangent to each other—or, as in the case of osculating curves, they meet with higher-order contact: not only sharing a common tangent (first order), but also matching curvature (second order) or even higher derivatives for a closer approximation. This point will be revisited in the following discussion.

To conclude this preliminary discussion, the oriented area enclosed by the two branches of the hysteresis loop spanned between $x_1<x<x_2$ 
\begin{align} 
A = \int_{x_1}^{x_2} (I_{\text{ascending}}(x) - I_{\text{descending}}(x)) \, dx 
\end{align}
can also be derived from the voltage-dependent susceptance
\begin{align}
    \int_{x_1}^{x_2} B(x) x \sqrt{1-x^2}dx.
\end{align}


\section{Self-Crossing Topology as a Phase Diagram for Hysteresis Loops}

Topological invariants—robust quantities that remain unchanged under continuous deformations—often appear as degrees, winding numbers, or signed intersection counts \cite{Whitney1937RegularClosedCurves,BurmanPolyak2011Whitney,GuilleminPollack1974}.  
Drawing from differential topology and singularity theory, we apply this robustness principle to the geometry of the hysteresis loop. Just as the sign of the discriminant of the characteristic equation cleanly partitions the parameter space of a damped harmonic oscillator into three, qualitatively distinct regimes (overdamped when $\Delta > 0$, critically damped when $\Delta = 0$, underdamped/oscillatory when $\Delta < 0$), the number of self-crossings (double points) of a generic smooth planar loop serves as a robust, integer-valued index \cite{Whitney1937RegularClosedCurves,GolubitskyGuillemin1973StableMappings}.

It remains strictly constant throughout open, connected regions of parameter space as long as the loop remains a smooth planar mapping with only transverse self-intersections (i.e., no self-tangencies and no cusp singularities) \cite{GolubitskyGuillemin1973StableMappings,Goryunov1998JPlusTheory}. A transverse self-intersection is a self-crossing where the two branches meet at a nonzero angle (their tangent directions at the crossing are distinct), i.e.\ an ``X''-type crossing rather than a tangential touch \cite{GuilleminPollack1974,BruceGiblin1992CurvesAndSingularities,EggersSuramlishvili2017}.

The crossing count changes only when the loop passes through a degenerate event---most importantly a self-tangency (where two branches touch without crossing) or a cusp singularity---or when, due to the restriction $x\in[-1,1]$, a previously existing self-intersection enters or exits the observed $x$-window (a truncation effect rather than a geometric singularity of the curve) \cite{BruceGiblin1992CurvesAndSingularities,GolubitskyGuillemin1973StableMappings}.

In this setting the index tracks qualitative shifts in the system's memory. The six-dimensional parameter space $(A, p, c_k)$ is therefore partitioned into large regions where the loop maintains a fixed number of transverse self-intersections, separated by thin bifurcation boundaries where the crossing count jumps through those non-generic events \cite{GolubitskyGuillemin1973StableMappings,Goryunov1998JPlusTheory}.

In the language of singularity theory, these regions are analogous to the structurally stable regimes familiar from the Zeeman catastrophe machine: once the control parameters lie inside one such region the qualitative behavior (number of crossings, hence the pattern of memory branches) is identical for all points inside it, no matter how far apart they are---exactly as the Zeeman machine exhibits one fixed folding pattern throughout each broad domain of its control plane \cite{Thom1975StructuralStabilityMorphogenesis,Zeeman1976CatastropheTheorySciAm,PostonStewart1978,Arnold1984Catastrophe}.
The thin separating boundaries are the catastrophe surfaces where the loop geometry undergoes a qualitative transition \cite{PostonStewart1978}.

To locate these boundaries we employ algebraic markers that detect the degeneracies in the polynomial. For a univariate polynomial constraint $P$ of degree $n$ with leading coefficient $a_n$, the discriminant vanishes precisely when $P$ develops a multiple root. This occurs when the resultant  of $P$ and its derivative $P'$ with respect to $x$ (i.e., $\operatorname{Res}(P, P')$)  vanishes, signaling a critical surface \cite{GKZ1994DiscriminantsResultants}:
\begin{equation}
\Delta(P) = \frac{(-1)^{n(n-1)/2}}{a_n} \operatorname{Res}(P, P') = 0.
\label{eq:DeltaResultant}
\end{equation}
More generally, the resultant of two polynomials detects shared roots, and resultants involving higher derivatives detect higher-order multiplicity conditions (higher-codimension degeneracies). This yields a rigorous algebraic stratification of the bifurcation set, classifying the stable regimes and the critical catastrophe surfaces that define the system's phase space in control parameters.

The classification is therefore as sharp and unambiguous as the discriminant classification of the harmonic oscillator or the catastrophe classification of the Zeeman machine: large, open regions of uniform qualitative behaviour, separated by explicitly computable algebraic surfaces \cite{EggersSuramlishvili2017,CoxLittleOShea2007}.

Having established the general role of the discriminant and resultant in detecting qualitative changes, we now turn to their concrete realization in our system. As shown in the previous sections, loop self-intersections occur precisely at the roots of the cubic susceptance $B(x)$. Since $B(x)$ has real coefficients, its discriminant $\Delta$ governs the root structure: $\Delta>0$ yields three distinct real roots, while $\Delta<0$ gives one real root and a complex-conjugate pair. The critical set where $\Delta=0$ corresponds to a multiple root and hence marks an algebraic degeneracy that can signal a bifurcation boundary.

However, three real roots do not necessarily imply three observable self-crossings of the hysteresis loop. A root contributes to the crossing count within a phase region only if it lies within the admissible sweep range $x\in[-1,1]$ (set by the driving amplitude via our normalization) and the corresponding intersection is transverse. Tangential self-contacts occur precisely on the bifurcation boundaries between regions and will be analysed in the following sections. These constraints will be incorporated into the region plots in subsequent steps.


With the structural context in place, we now address a simple but revealing question: how strongly does the nonlinear perturbation in \eqref{eq:IVscaled} act on the baseline \(G=1\)? Quantifying this influence provides both a measure of nonlinearity and a means to simplify the partitioning of the 6-dimensional parameter space. To this end, we evaluate the maximum of each term in \eqref{eq:IVscaled} over the unit circle, since one period of the sinusoidal drive traces that circle.

The maximum value of the first-order nonlinear term 
$
\left| -x + p y \right|
$
on the unit circle is 
\[
\text{M}_1 = \sqrt{1 + p^2}.
\] 
Consequently, it is convenient to reparameterize the \(c_1\)-term in \eqref{eq:IVscaled}, $A c_1(x - p y)/(1 + p^2) $, using a new variable \(\xi_1\), so that \(\xi_1\)—rather than \(c_1\)—is compared directly to the unit conductance \(G=1\)
\begin{align}
c_1 =\xi_1 \left(A \frac{\, \text{M}_1}{1 + p^2} \right)^{-1}.
\end{align}
For the second-order term 
$
\left| 2 p^2 + x^2 - 2 p x y \right|,
$
the maximum on the unit circle is 
\[
\text{M}_2 = \tfrac{1}{2}\,\bigl( 1 + 4p^2 + \sqrt{\,1 + 4p^2\,} \bigr).
\]
The formulas for the third and fourth order maxima are too lengthy. Hence, we will denote them as $\text{M}_3$ and $\text{M}_4$. This normalization over the unit circle is particular to the sinusoidal input, whereas the initial normalization \eqref{ScaledGeneratingFunction} holds in general input wave forms.


Expressed in the normalized \(\xi\) variables, the susceptance \(B\) reads
\begin{align}
B=p \Bigg( &-\frac{\xi_1}{\text{M}_1} 
- \frac{2 x \, \xi_2}{\text{M}_2} 
- \frac{3 (x^2 + p^2 (2 + x^2)) \, \xi_3}{\text{M}_3} \nonumber \\
&- \frac{4 x (x^2 + p^2 (6 + 4 x^2)) \, \xi_4}{\text{M}_4} \Bigg).
\end{align}

We absorb the input amplitude \(A\) into the four parameters \(\xi_k\), leaving only the frequency \(p\) as an extrinsic parameter. Consequently, after analyzing the crossings in terms of \((\xi_1,\dots,\xi_4,p)\), the influence of \(A\) is conveyed solely through how \(\xi_k\) varies with \(A\). 

Throughout the paper, we assume $\xi_4 \neq 0$ to ensure that the admittance is quartic in the voltage. 
For this device, formula (\ref{eq:DeltaResultant}) reduces to
\begin{equation}\label{eq:DeltaDevice}
    \Delta = \frac{M_4}{16} \, \xi_4^{-1} \, \operatorname{Res}(B, B').
\end{equation}
Because \( M_4 > 0 \), the sign of \(\Delta\) coincides with the sign of \(\xi_4^{-1} \operatorname{Res}(B, B')\). Thus, the bifurcation regions are completely determined by the susceptance \( B(x) \).


Having established the relevant variables, we now adopt a cusp-catastrophe framework to examine the bifurcations of hysteresis crossings as the parameters change. This choice is motivated by the broader context of catastrophe theory, which offers a unified classification of systems exhibiting abrupt state transitions—suggesting that other classes of memristors may correspond to different catastrophe types.

\section{From Conductance–Susceptance Relation to Cusp Catastrophe: Control Parameters and Crossing Bifurcations}

A particularly advantageous analytic relationship holds between the nonlinear susceptance $B(x)$ and the nonlinear conductance $G(x)$

\begin{align}
    B(x)=-p\frac{\partial G(x)}{\partial x}
\end{align}

\begin{figure}[h]
\includegraphics[width=0.40\textwidth]{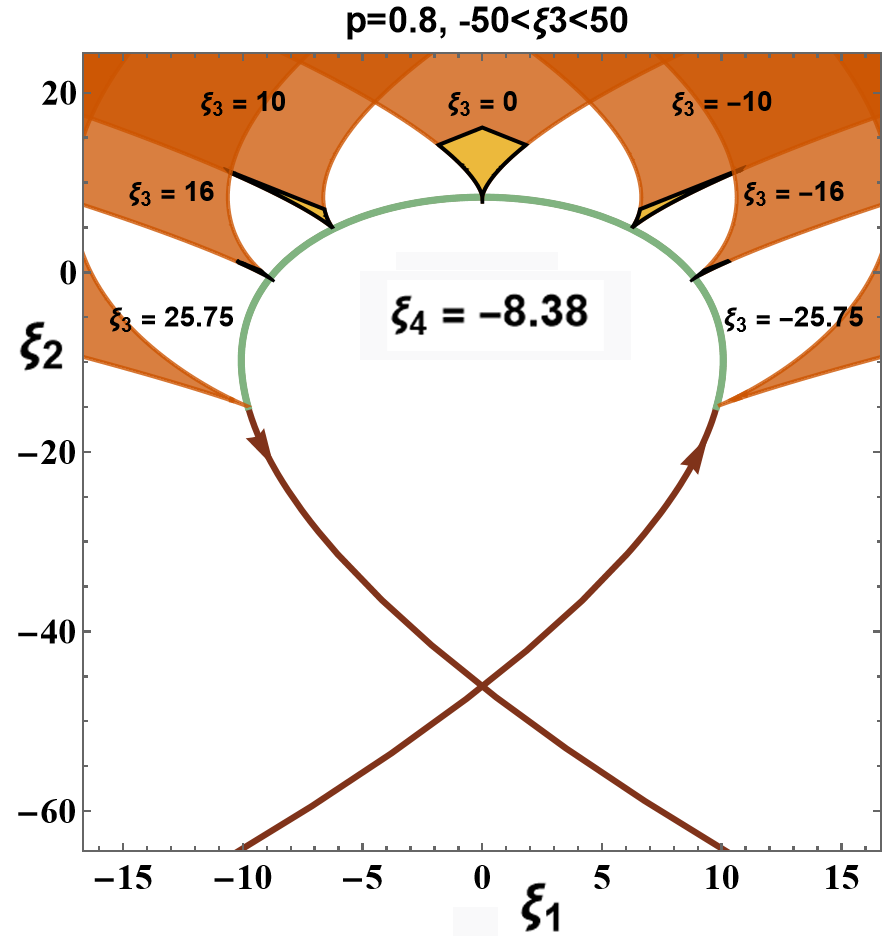}
\caption{\label{fig:RegionCrossingTrajectoryV4}Illustration of the cusp region (orange) and the subregion with four crossing points (yellow) in the $(\xi_1, \xi_2)$ plane for fixed $\xi_3$ and $\xi_4$.
    }
\end{figure}

This allows interpreting \(B(x)=-\partial_x U\) for the potential $U(x)=p\, G(x)$ in the spirit of the cusp equilibrium condition, with \(x\) as the state variable \cite{ Thom1975StructuralStabilityMorphogenesis,GKZ1994DiscriminantsResultants,Zeeman1976CatastropheTheorySciAm}. We now take \((\xi_1, \xi_2)\) as the control parameters for the present analysis, treating \((\xi_3, \xi_4)\) as auxiliary labels that parameterize the model family. Different pairs will be selected as control parameters in subsequent sections.
We therefore focus on the cusp region (\(\Delta>0\)), where \(\Delta\) denotes the discriminant of \(B\) with respect to \(x\); here the cubic equation \(B(x)=0\) has three real solutions, and the hysteresis loop may exhibit the maximal number of extra crossings. To differentiate the inherent zero-voltage crossing, defined by the model itself, from all other crossings, we refer to the latter as extra or additional crossings. Figure \ref{fig:RegionCrossingTrajectoryV4} illustrates seven distinct cusp regions, with their positions determined by the variable $\xi_3$ when $\xi_4$ and $p$ are held constant.

The cusp catastrophe allows sudden jumps between phases as \(\xi_1\) or \(\xi_2\) crosses the bifurcation boundary \(\Delta=0\). For \(\Delta<0\), the cubic \(B(x)\) has a single real root—i.e., the susceptance is zero at a single point—so the hysteresis loop can exhibit at most one additional crossing. In the control parameters $(\xi_1, \xi_2)$, the cusp region is generated by the monomials $\xi_2^3$ and $\xi_2 \xi_1$ appearing in the sign-determining factor of (\ref{eq:DeltaDevice})
\begin{align}\label{eq:Delta}
\xi_4^{-1}\operatorname{Res}(B, B')=& -\frac{512 p^5 (1 + 4 p^2)^2 \, \xi_4}{\text{M}_2^3 \, \text{M}_4^2}\xi_2^3+\\\nonumber
&\frac{1728 p^5 (1 + p^2)(1 + 4 p^2)^2 \, \xi_3 \, \xi_4}{\text{M}_1 \, \text{M}_2 \, \text{M}_3 \, \text{M}_4^2}\xi_1\xi_2+\cdots
\end{align}
The terms less critical to the cusp region are represented by the ellipsis.
Equation \eqref{eq:Delta} does not immediately exhibit the semicubical parabola form. In catastrophe theory, appropriate coordinate transformations are required to reveal such canonical structures, aligning the system with the universal form defined by Thom’s classification theorem.

Nonetheless, adopting an alternative coordinate system does not diminish the validity or significance of the resulting insights. We therefore keep \((\xi_1,\xi_2)\) as control parameters, as they carry direct physical meaning for the device under study. This choice clarifies how we traverse the four-dimensional space of the generating function—by fixing \((\xi_3,\xi_4)\) and and exploring \((\xi_1,\xi_2)\) —and sets the framework for the analysis that follows.

\section{Bifurcation Set of the Cusp Catastrophe: Structure and Significance of the Cusp Point}

The bifurcation set for the cusp catastrophe is defined by the equation \( \Delta = 0 \) \cite{Arnold1984Catastrophe}. This set forms the boundary of the cusp region in the control parameter space $(\xi_1,\xi_2)$, and it consists of two curves (the fold curves) that intersect at a single point. This intersection point is a cusp. This point is where the system transitions from having one equilibrium to three equilibria (or vice versa) as the control parameters cross the bifurcation set. This point is a critical singularity where the system’s behavior is highly sensitive to changes in the control parameters, marking the apex of the cusp-shaped region where bistability occurs.

Fixing \((\xi_3,\xi_4)\), the cusp location in the \((\xi_1,\xi_2)\) plane is specified by \eqref{eq:Xi1AtCusp} and \eqref{eq:Xi2AtCusp}.
\begin{align}\label{eq:Xi1AtCusp}
&\xi_1=\frac{\text{M}_1 \, \xi_3}{16 \, \text{M}_3^3 \left( \xi_4 + 4 p^2 \xi_4 \right)^2}\\\nonumber
&(\text{M}_4^2 (1 + p^2)^3 \xi_3^2 - 96 \, \text{M}_3^2 p^2 (1 + 4 p^2)^2 \xi_4^2)
\end{align}
\begin{align}\label{eq:Xi2AtCusp}
\xi_2 = -\frac{12 \, \text{M}_2 \, p^2 \, \xi_4}{\text{M}_4} + \frac{3 \, \text{M}_2 \, \text{M}_4 (1 + p^2)^2 \, \xi_3^2}{8 \, \text{M}_3^2 (\xi_4 + 4 \, p^2 \, \xi_4)}
\end{align}

At the critical point, the susceptance forms a perfect cubic  function with three degenerate voltage roots
\begin{align}\label{BatTheCuspPoint}
 B_{\text{cusp}}(x) =  -\frac{p \left( \text{M}_4 (1 + p^2) \xi_3 + 4 \, \text{M}_3 (1 + 4 p^2) x \, \xi_4 \right)^3}{16 \, \text{M}_3^3 \, \text{M}_4 \left( \xi_4 + 4 p^2 \xi_4 \right)^2}
\end{align}

At this stage, we have explicit expressions for both the cusp region and the cusp point. The cusp region gives a necessary condition for the hysteresis curve to exhibit up to three additional crossings, beyond the trivial crossing at zero voltage.

For these roots to produce crossings in the scaled $I\!-\!V$ plane, however, they must lie within the interval $-1 \le x \le 1$. This requirement imposes further constraints, namely $B(\pm 1)=0$. In the $(\xi_1,\xi_2)$ plane, these conditions correspond to two boundary straight lines.

Within the cusp region, these two lines delineate a subregion (marked yellow in Fig.\ref{fig:RegionCrossingTrajectoryV4}) of the \((\xi_1,\xi_2)\)-plane that supports three additional crossing points. For fixed values of $\xi_4=-8.38$ and $p=0.8$, the yellow region progressively narrows as $\xi_3$ changes from 0 to $\pm 25.75$, vanishing at $\xi_3=\pm 25.75$. At these limiting values, the points  $x = \pm1$ become triple roots, marking the onset of additional crossings at the hysteresis turning points. The coincidence of these crossings with the turning points identifies the range of $\xi_3$ over which a hysteresis with three additional crossings (or four total crossings) exists

\begin{align}\label{SegmentOfXi3}
    |\xi_3| < \left| \frac{4 {M_3} (1 + 4 p^2)}{{M_4} (1 + p^2)} \xi_4 \right|
\end{align}

In Figure \ref{fig:RegionCrossingTrajectoryV4},  as $|\xi_3|$ increases, the cusp point traces a trajectory defined by  (\ref{eq:Xi1AtCusp}) and (\ref{eq:Xi2AtCusp}).
The sage-green segment of this curve denotes the range of validity from (\ref{SegmentOfXi3}), beyond which the four-crossing subregion ceases to exist. As $\xi_3$ approaches the boundaries of  (\ref{SegmentOfXi3}), the four-crossing subregions become increasingly asymmetric and progressively narrower. At these boundaries, one crossing exits the hysteresis loop through its endpoints.

 \section{Phase Analysis of Hysteresis Loops with Multiple Crossings}

\begin{figure}[h]
\includegraphics[width=0.35\textwidth]{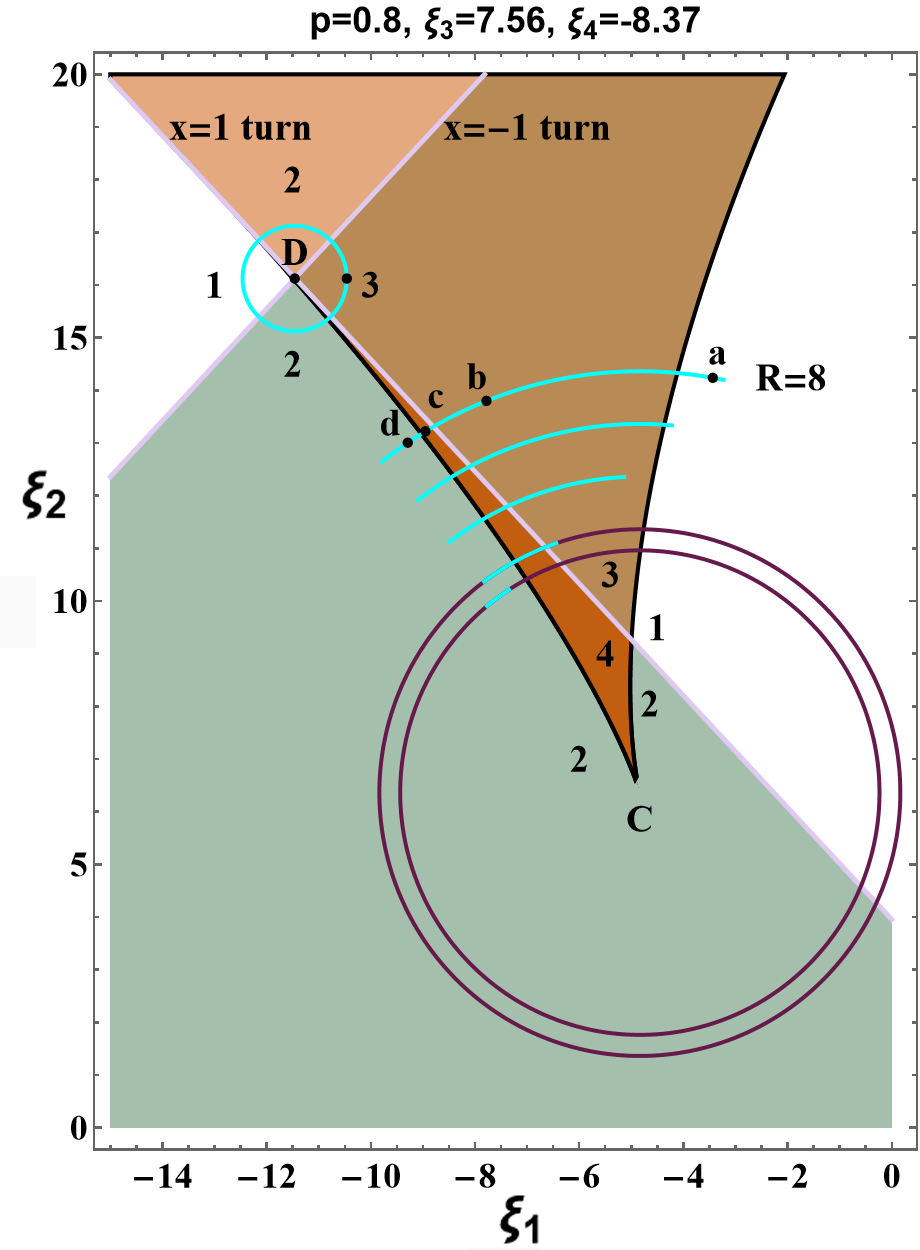}
    \caption{\label{fig:CircleTheCuspV1} Phases for hysteresis loops with different crossing numbers. The regions are labeled with the total crossing number, including the instance at zero voltage. The cusp is located at point C $  (\xi_1 = -4.84, \xi_2 = 6.36)  $, and point D is at $  (\xi_1 = -11.46, \xi_2 = 16.12)  $. Circles of radii $  R = 4.6, 5, 6, 7, 8  $ are centered at C, and a circle of radius $  R = 1  $ is centered at D. The cyan arcs on the circles centered at C indicate the range of $  (\xi_1, \xi_2)  $ values for which the device exhibits passive behavior \eqref{PassiveDevice}. }
\end{figure}

Building upon the phase diagram presented in Fig.~\ref{fig:RegionCrossingTrajectoryV4}, we consider the specific case \(\xi_3 = 7.56\) and  \(\xi_4 = -8.37\), chosen to illustrate the application of the cusp region as depicted in Fig.~\ref{fig:CircleTheCuspV1}. The cyan arcs on the circles centered at point C represent the range of $(\xi_1, \xi_2)$ values for which the device exhibits passive behavior \eqref{PassiveDevice}. For the circle of smallest radius, the cyan arc is relatively short and lies entirely within the phase region associated with two total crossings. On the next circle, the arc transitions from the phase corresponding to three total crossings to four, and finally terminates in the two-crossing region. As the radius grows, the arcs extend over a broader range of phases. 

\begin{figure}[h]
\includegraphics[width=0.475\textwidth]{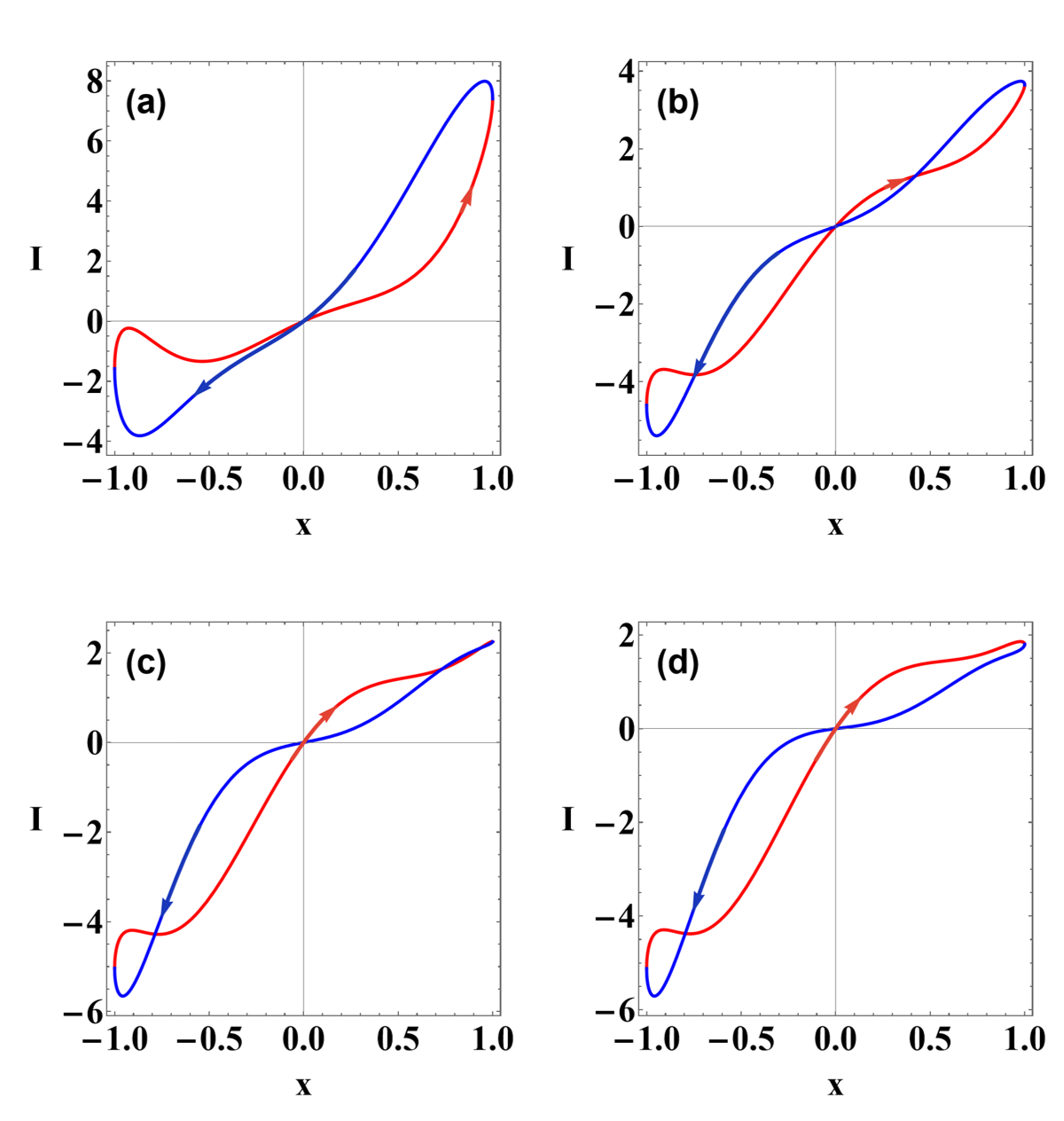}
\captionsetup{justification=raggedright, singlelinecheck=false}
\caption{\label{fig:AroundPointC}
The hysteresis loops associated with the four black dots labeled a--d on the circle of radius $R=8$ from Fig.~\ref{fig:CircleTheCuspV1}. Panels~(a)--(d) show the loops for points a--d, respectively, with $(\xi_1,\xi_2)=(-3.43,14.23)$, $(-7.78,13.80)$, $(-8.95,13.22)$, $(-9.29,13.00)$.}
\end{figure}

In the case of $R = 8$, the arc originates at the single-crossing point (zero voltage), traverses the regions with three and four total crossings, and terminates in the two-crossing phase. 
The hysteresis loops corresponding to the four selected points labeled a--d along this trajectory are shown in Fig.~\ref{fig:AroundPointC}. When the total number of crossings changes by two—as is the case from panel (c) to panel (d)—two crossings in (c) move toward each other, merge into one, and then, as seen in (d), the branches separate again. In the $(\xi_1, \xi_2)$-plane, this phenomenon is observed when crossing the left branch of the bifurcation set from the four-crossing region to the two-crossing region, or the right branch from the one-crossing region to the three-crossing region.

The annihilation of two crossing points is observed in the complex plane of the susceptance roots when two real roots converge, collide, and subsequently both diverge into the complex plane, moving off the real axis as their imaginary parts become nonzero while their real parts remain within the interval $[-1, 1]$. Conversely, the phenomenon described corresponds to the simultaneous creation of two crossing points. 

The lines $B(\pm 1)=0$, referred to as the $x = \pm 1$ turns, delineate transitions where a crossing point either enters or exits the hysteresis loop through its endpoints upon crossing these lines. The influence of the line $x = 1$ is visible in the hysteresis loop in Fig.\ref{fig:AroundPointC}(c) because the corresponding point in Fig.\ref{fig:CircleTheCuspV1} lies near this line. During the transition from panel (b) to panel (c), the fourth crossing enters the hysteresis loop from its rightmost turning point at $x = 1$.

\begin{figure}[h]
\includegraphics[width=0.475\textwidth]{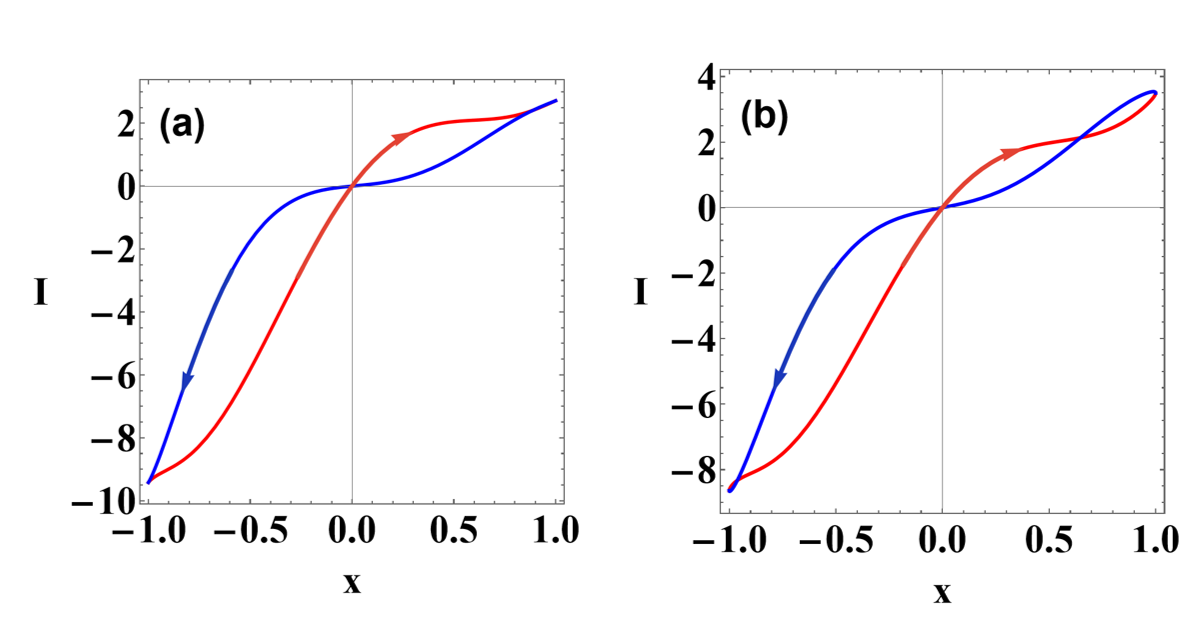}
\captionsetup{justification=raggedright, singlelinecheck=false}
\caption{\label{fig:AroundPointD}
Hysteresis loops corresponding to point D from Fig.~\ref{fig:CircleTheCuspV1}. Panel (a) shows the loop at point D with $(\xi_1 = -11.46, \xi_2 = 16.12)$, while panel (b) corresponds to the nearby point on the $R = 1$ circle around D, located in the region with three crossings at $(\xi_1 = -10.46, \xi_2 = 16.12)$.
}
\end{figure}

\begin{figure}[h]
\includegraphics[width=0.475\textwidth]{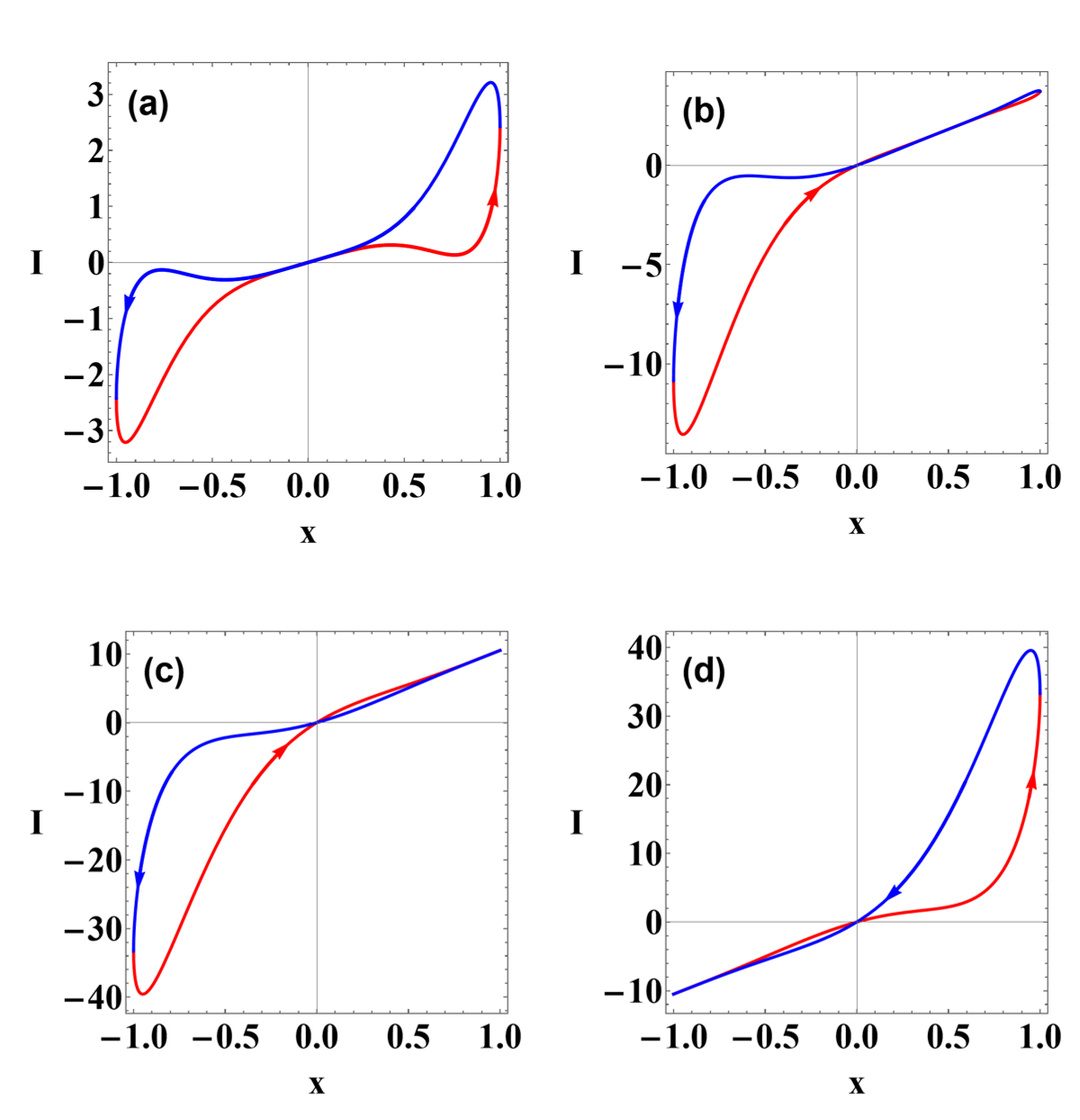}
\captionsetup{justification=raggedright, singlelinecheck=false}
\caption{\label{fig:AtCusp}
The hysteresis loops associated with different cusps . For panel (a) the triple root to panel (d) the cusps are different with the pairs $\xi_4=8.73$ and $\xi_3$ given by $0$,$-12.86$, $-25.72$, and $25.72$, all for $p=0.8$.  
}
\end{figure}

The disappearance of a crossing points at the extremities of the hysteresis loop is observed in the complex plane of the susceptance roots as a single real root moves away from either $-1$ or $+1$ along the real axis.
Near point D in Fig.~\ref{fig:CircleTheCuspV1}, the full circular trajectory exhibits passive behavior. Since this circle intersects both lines $x = \pm 1$, certain crossing points appear and disappear through each of the hysteresis turning points. Figure~\ref{fig:AroundPointD}(a) shows two crossings emerging from the turning points at $x = \pm 1$, where the hysteresis branches become tangent at their outermost points. In Fig.~\ref{fig:AroundPointD}(b), the two crossings evolve in opposite directions along the hysteresis branches, with one descending and the other ascending relative to panel (a).
The cusp  C in Fig. \ref{fig:AroundPointC} is unattainable for a passive device, as condition \eqref{PassiveDevice} is not satisfied. However, other cusp points are attainable, and their corresponding hysteresis behaviors are shown in Fig.\ref{fig:AtCusp}.  

The notable feature at the cusp is that the crossing becomes triply degenerate—or even quartically degenerate if the additional three crossings occur at zero voltage. Extracting from (\ref{BatTheCuspPoint}) the cusp-crossing relation 
\begin{align}\label{eq:xAtCusps}
    x_{\text{cusp}} = -\frac{M_4 (1 + p^2) \xi_{3}}{4 M_3 (1 + 4 p^2) \xi_{4}}
\end{align} we positioned the cusp crossing at various prescribed voltages. In Fig. \ref{fig:AtCusp}(a), the crossing at zero voltage is quartically degenerate. In panel (b), the triply degenerate cusp crossing is located at  $x=0.5$. In panels (c) and (d), the triple crossing is positioned at 
$x=1$ and 
$x=-1$, respectively, producing hysteresis loops that are asymmetric between quadrants I and III and, moreover, exhibit different orientations.

Before concluding this section, we note the utility of phase maps such as Fig.\ref{fig:CircleTheCuspV1}. Interpreting Figs.\ref{fig:AroundPointC} and \ref{fig:AroundPointD} within such a phase map makes it straightforward to infer the associated hysteresis loop shape from the parameter point’s position—whether deep into a phase region or near a cusp, an $x=\pm1$-line, or a bifurcation curve.

\section{Role of the Cubic Voltage Dependence of the Nonlinear Admittance in Three-Dimensional Analysis}

A strong motivation for extending the control parameters to include $\xi_3$ is the linear dependence of the hysteresis loop orientation (\ref{eq:B0}) on $\xi_1$ and $\xi_3$, given explicitly by the following formula:

\begin{equation}
B(0)=-\frac{p ( M_3 \xi_1 + 6 M_1 p^2 \xi_3 )}{M_1 M_3}.
\end{equation}

The orientation of the loop, determined by the sign of $B(0)$, further underscores the importance of $\xi_3$. To fully visualize the loop's orientation in relation to additional crossings, it is necessary to consider the device not only in the two-dimensional plane defined by $(\xi_1, \xi_2)$, but also within a three-dimensional parameter space $(\xi_1, \xi_2, \xi_3)$, while treating $\xi_4$ as a separate, non-zero, one-dimensional parameter. 

\begin{figure}[h]
\includegraphics[width=0.35\textwidth]{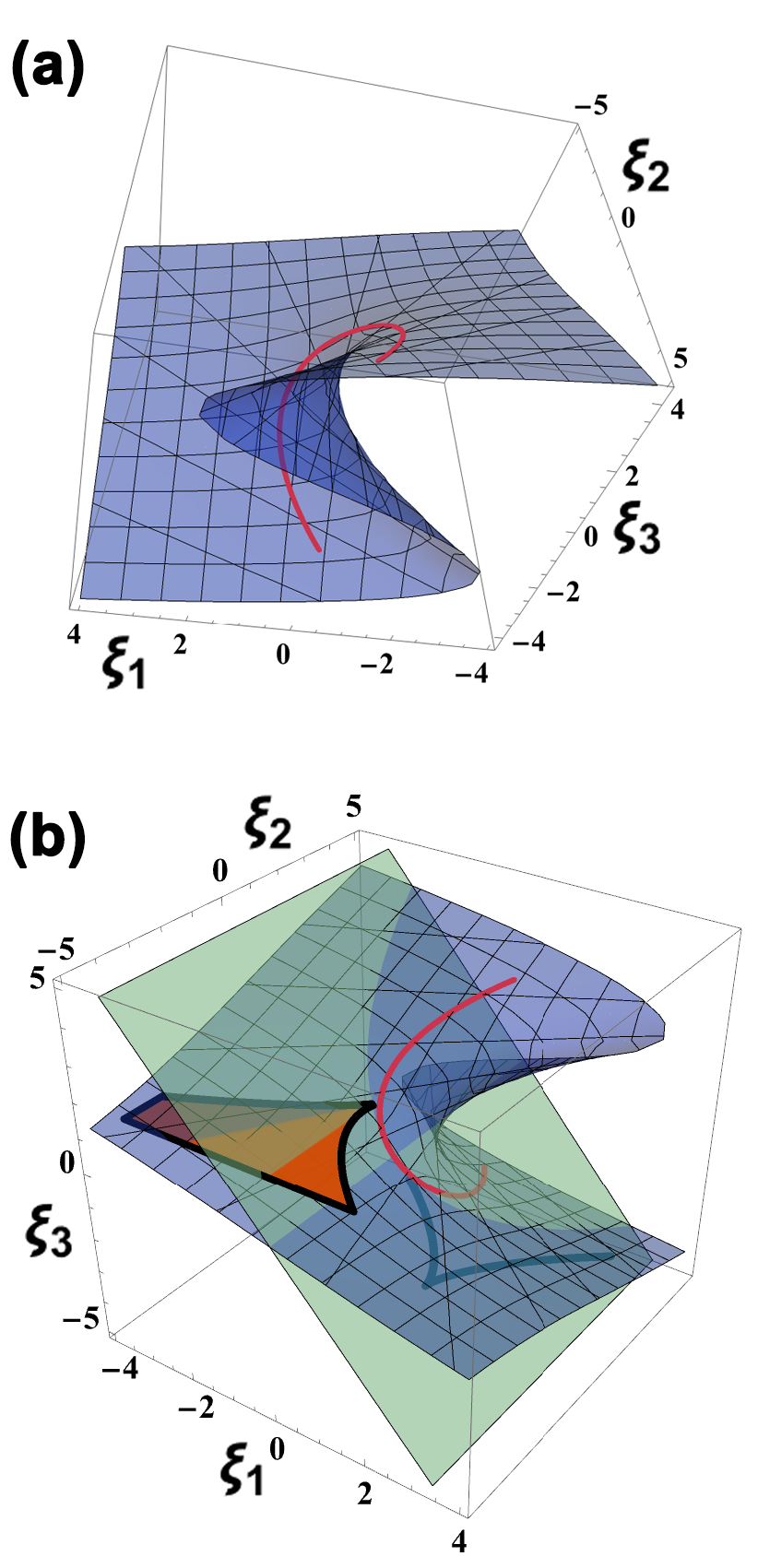}
\caption{\label{fig:BlueSurfaceSpaceXi}\raggedright
Two complementary views in the three-dimensional parameter space $  (\xi_1, \xi_2, \xi_3)  $. Panel (a) shows a simplified front view containing only the blue folded surface and red cusp curve so that the fold can be clearly identified.(b) Rotated view from the opposite side with additional structures: an orange cusp region bounded by black bifurcation curves, a light-green plane traversed when the hysteresis loop orientation changes, and the dark-green projection of the red curve onto the $(\xi_1, \xi_2)$-plane.}
\end{figure}

In this three-dimensional space, the change in the orientation of the hysteresis loop occurs as the light-green plane in Fig.~\ref{fig:BlueSurfaceSpaceXi}(b) is traversed.

Another advantage of the three-dimensional parameter space is that the cusp, defined by (\ref{eq:Xi1AtCusp}) and (\ref{eq:Xi2AtCusp}), no longer appears as an isolated point but instead traces a curve, shown in red and, for clarity, displayed from two distinct viewpoints in panels (a) and (b) of Fig.~\ref{fig:BlueSurfaceSpaceXi}. This, in turn, enables a more systematic exploration of hysteresis loops in the vicinity of the cusp curve within the three-dimensional parameter space. In this exploration, we move away from the cusp by relaxing the third-order degeneracy condition associated with the crossing in (\ref{eq:xAtCusps}), and instead require only that 
$B(x)$ still exhibits a crossing at (\ref{eq:xAtCusps}) without imposing any degeneracy.

This relaxed condition is satisfied by all parameter triples $(\xi_1, \xi_2, \xi_3)$ lying on a surface within the same resultant family that contains the discriminant. Specifically, this surface is defined by the resultant versus $x$ of the susceptance $B(x)$ and its second derivative with respect to $x$. Recall that the discriminant itself is based on the  resultant versus $x$ of $B(x)$ and its first derivative, making this construction a natural extension. 

The blue surface in both panels (a) and (b) of Fig.~\ref{fig:BlueSurfaceSpaceXi} is the one defined above and contains the red curve that represents the locus of cusp points. The blue surface is a folded sheet embedded in the three-dimensional space
\((\xi_1,\xi_2,\xi_3)\).In panel~(a), the surface is viewed from the front,
so the fold can be seen as the edge of the sheet bending inward, with its
fold most clearly visible near the plane \(\xi_2=5\). 
In panel~(b), the entire configuration has been rotated, showing the fold from the opposite side and additional structures.
One such structure, in orange, is an example of a cusp region. Within the three-dimensional space $  (\xi_1, \xi_2, \xi_3)  $, varying $  \xi_3  $ translates the entire cusp region parallel to the $  (\xi_1, \xi_2)  $-plane.
As mentioned above, when this occurs, the cusp point traces the three-dimensional parametric curve (red), which, when projected onto the plane $  (\xi_1, \xi_2)  $, yields the green curve.

Before leaving Fig.~\ref{fig:BlueSurfaceSpaceXi}, we note that the blue folded surface represents the equilibrium surface of a cusp catastrophe, with \(\xi_3\) as the state variable and \((\xi_1,\xi_2)\) as the control parameters. This allows us to identify the cusp region and its bifurcation set, represented by the dark-green curve in the \((\xi_1,\xi_2)\)-plane in Figs.~\ref{fig:BlueSurfaceSpaceXi}, which is independent of \(\xi_3\).

The parametric equations for this green bifurcation set, with parameter \(s\), are given by
\begin{align}
\xi_1(s) &= 2 s^3 M_1 \\
\xi_2(s) &= s^2 \frac{3 (1 + 4p^2)^{1/3} M_2 \xi_4^{1/3}}{M_4^{1/3}} + \frac{36 p^4 M_2 \xi_4}{(1 + p^2) M_4}.
\end{align}

To avoid confusion, the family of bifurcation sets associated with the orange cusp regions—which depend on \(\xi_3\)—is obtained by treating the variable \(x\) (rather than \(\xi_3\)) as the state variable.

\begin{figure}[h]
\includegraphics[width=0.45\textwidth]{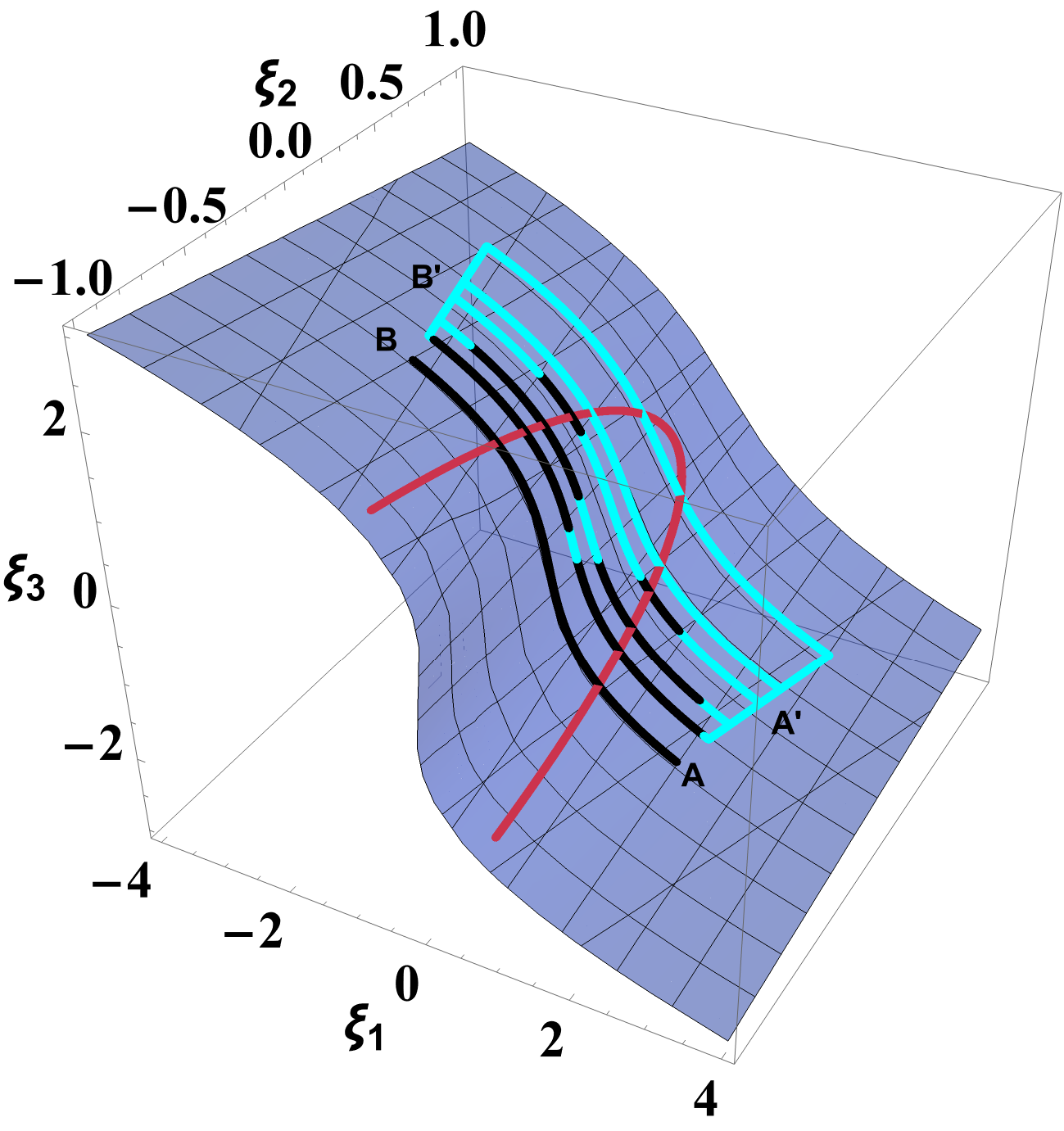}
\caption{\label{fig:Jumps}
Structure of the three-dimensional parameter space \((\xi_1,\xi_2,\xi_3)\) for \(\xi_4 = -0.6\). The red curve marks the locus of cusp points lying on the blue folded surface. Shown are constant-\(\xi_2\) sections, starting from the AB-path at \(\xi_2 = -0.25\) (none of whose points represent a passive device) and continuing to the section at \(\xi_2 = -0.1\). Path segments corresponding to passive devices are highlighted in cyan. The cusp point traces the red curve on the blue surface associated with the second resultant.}
\end{figure}

On a related note, the fold surface of the cusp catastrophe is mainly invoked to illustrate the abrupt state transitions a system undergoes when driven across regions of rapid variation \cite{Zeeman1976CatastropheTheorySciAm}. In our device, we find that abrupt transitions also occur on this fold surface once we recognize that the relevant notion of ‘state’ pertains to whether the device operates passively or actively. In this context, the state is effectively binary: allowable when the device remains passive (\ref{PassiveDevice}), and forbidden when the hysteresis loop enters quadrants II or IV, indicating active behavior.
Figure~\ref{fig:Jumps} illustrates this behavior. Each curve is a cross-section at fixed \(\xi_2\), with \(\xi_3\) varying from \(-2\) to \(2\). The first curve, at \(\xi_2 = -0.25\), corresponds entirely to active behavior; all points are shown in black to denote forbidden states. At \(\xi_2 = -0.1\), the endpoints of the path appear in cyan, indicating the emergence of allowable (passive) states. As \(\xi_2\) increases through \(0,\, 0.15,\, 0.25,\) and \(0.50\), the allowed and forbidden segments alternate along each path, with the forbidden interval shrinking progressively and eventually disappearing for \(\xi_2 \ge 0.25\).

Particularly informative paths occur at \(\xi_2 = 0\) and \(\xi_2 = 0.15\). 
As \(\xi_3\) is increased from \(-2\) to \(2\), maintaining the device along one of these paths may force a sudden transition through the forbidden region. In such cases, the passive-device model ceases to apply within the forbidden zone; however, the physical device may undergo a rapid transition that effectively “skips over’’ this region and re-enters an allowable state described once again by the model. 

This reasoning parallels cusp catastrophe theory, which classifies the equilibrium branches (stable and unstable) and predicts whether a system follows them smoothly or undergoes discontinuous jumps as the control parameters vary. However, catastrophe theory itself does not describe the time-resolved dynamics or the mechanisms of the transition.

\section{Amplitude--Frequency Flow in Tschirnhaus Canonical Cusp Coordinates: Fixed Cusp Region and Moving Boundaries}

The final section is devoted to visualizing how changes in amplitude and frequency generate a flow in phase space. Here, the device's internal parameters $c_k$, which define the generating function, are held fixed, while the amplitude and frequency are treated as control variables. Recall that the amplitude is absorbed into the variables $\xi_k$, and that the corresponding maxima $M_k$ depend on the frequency $p$.

In this section we move beyond the $\xi_k$ variables, which are directly tied to the nonlinear contribution to the admittance. This shift is motivated not only by the need to treat amplitude explicitly, but also by the fact that the phase-space structure is now familiar, allowing us to introduce a change of variables that maps any cusp boundary into a canonical form. In particular, to reduce a general cubic to the simpler depressed form, we use the standard Tschirnhaus transformation---a shift of the unknown that eliminates the quadratic term---so that the subsequent analysis can be carried out in terms of the resulting canonical coefficients \cite{Arnold1984Catastrophe,vanDerWaerden2003Algebra}.

We adopt the canonical coordinates \((u, w)\) such that the cusp condition takes the standard form
\[
w^3 - u^2 = \xi_4^{-1} \operatorname{Res}(B, B')\,.
\]

\begin{figure}[h]
\includegraphics[width=0.48\textwidth]{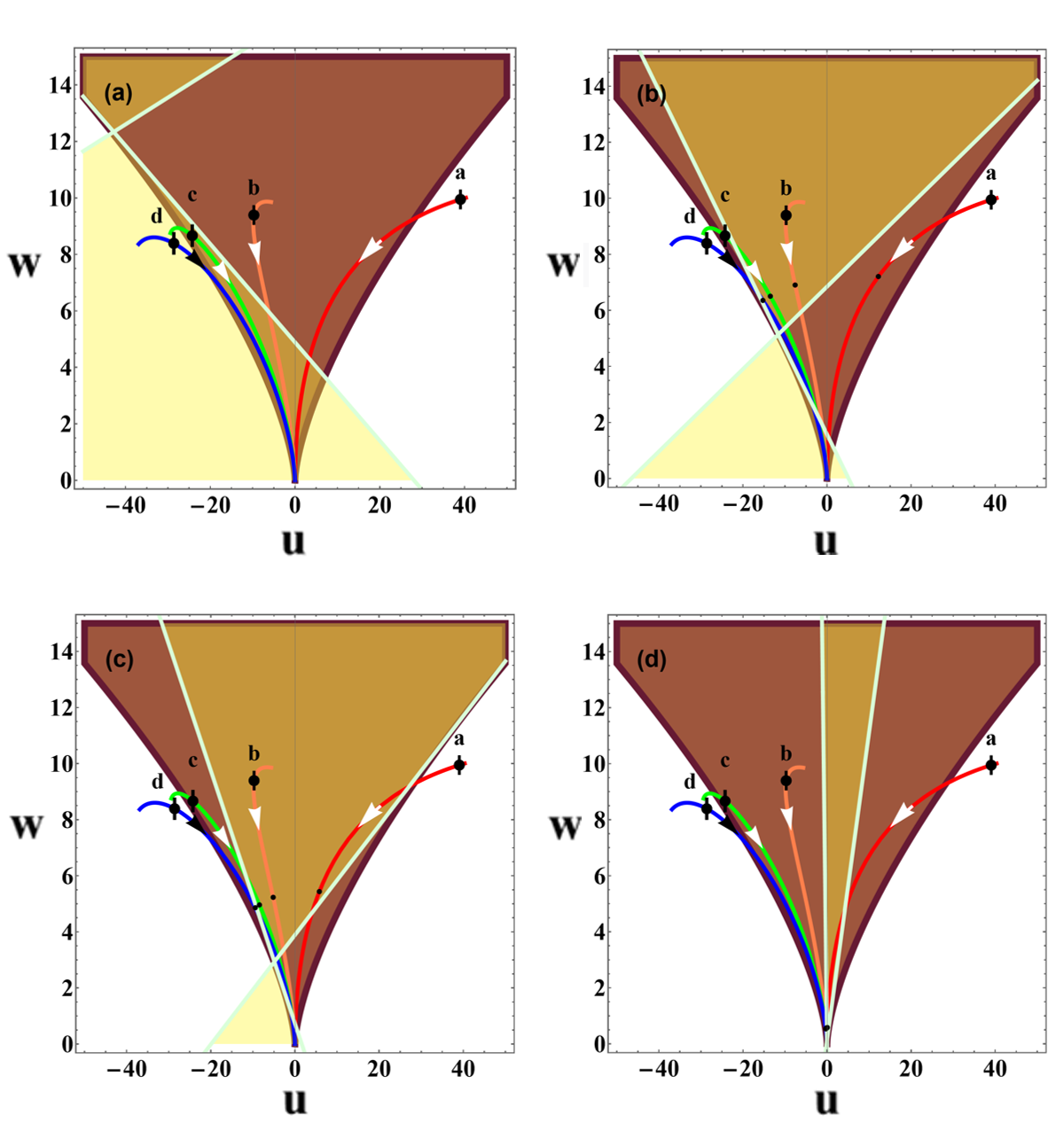}
\caption{\label{fig:FlowAmplitudes}
Trajectories in the $(u,w)$ phase plane generated by varying the amplitude $A$ for the points labeled a--d in Fig.~\ref{fig:CircleTheCuspV1} (shown here with the same labels). In each panel, the initial location at $A=1$ is indicated by a short vertical tick. From panel (a) to (d), the amplitude takes the values $A=1$, $0.835$, $0.742$, and $0.350$; the corresponding positions reached along each trajectory are marked. The cusp region remains fixed, whereas the straight-line boundaries defined by $B(\pm 1)=0$ shift with the amplitude.
}
\end{figure}

\begin{figure}[h]
\includegraphics[width=0.48\textwidth]{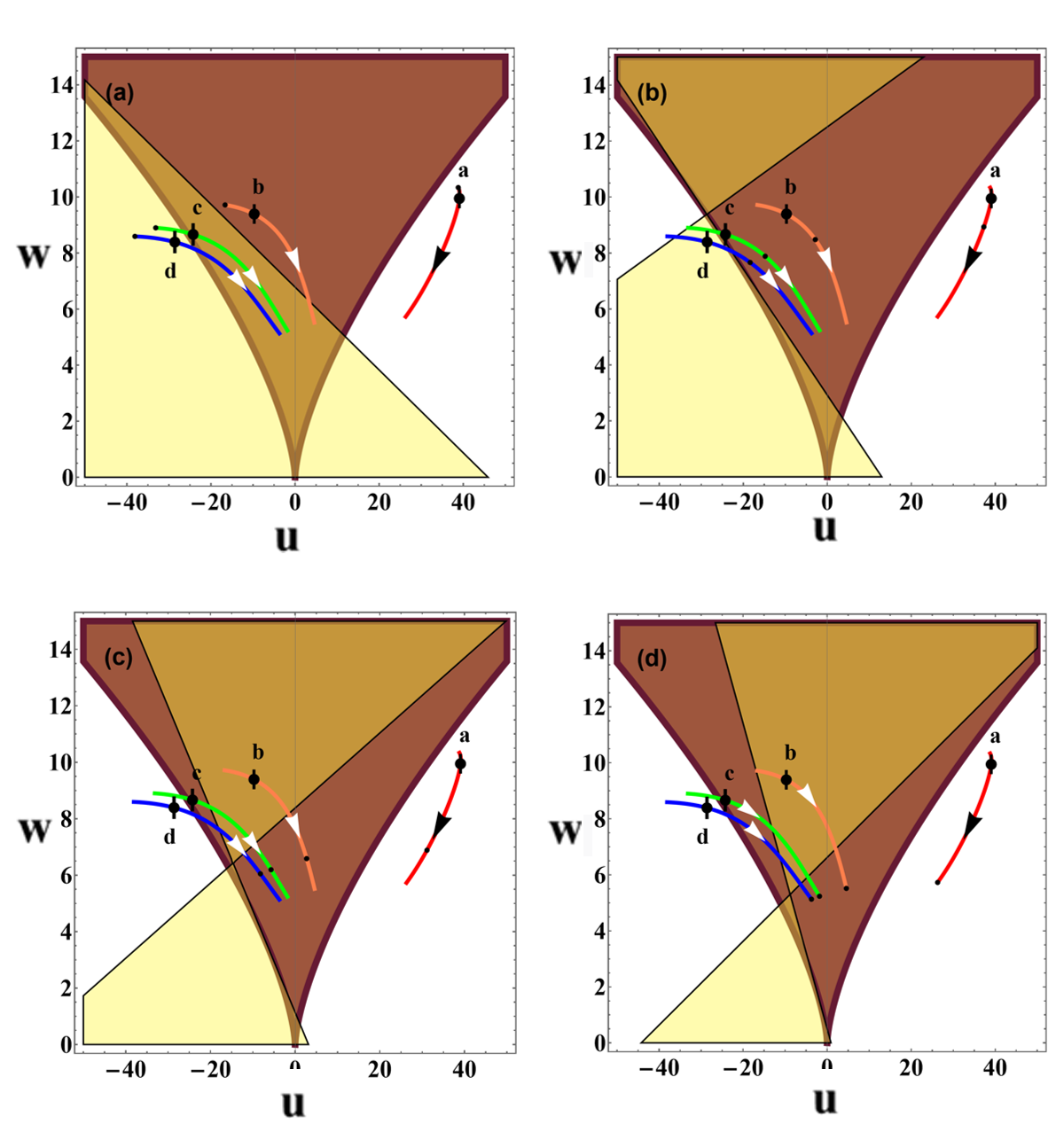}
    \caption{\label{fig:FlowFrequency} Trajectories in the $(u,w)$ phase plane generated by varying the frequency $p$ for the points labeled a--d in Fig.~\ref{fig:CircleTheCuspV1} (shown here with the same labels).  In each panel, the initial location at $p=0.8$ is indicated by a short vertical tick. From panel (a) to (d), the amplitude takes the values $p=0.7$, $1.0$, $1.5$, and $2.0$; the corresponding positions reached along each trajectory are marked. The cusp region remains fixed, whereas the straight-line boundaries defined by $B(\pm 1)=0$ shift with the amplitude.
    }
\end{figure}

The transformation from $(\xi_1,\ldots,\xi_4)$ and the frequency $p$ to $(u,w)$ is linear in $(\xi_1,\xi_2)$, but nonlinear in $(\xi_3,\xi_4,p)$. This linear dependence on $(\xi_1,\xi_2)$ enables a direct mapping of the phase portrait studied in the $(\xi_1,\xi_2)$ plane (Fig.~\ref{fig:CircleTheCuspV1}) to the corresponding phase diagram in canonical coordinates (Figs.~\ref{fig:FlowAmplitudes}--\ref{fig:FlowFrequency}). A key advantage of the $(u,w)$ parametrization is that the cusp region is fixed and does not vary with amplitude or frequency. By contrast, the boundary constraints $B(\pm 1)=0$ shift with changes in amplitude and frequency. Nevertheless, since $(u,w)$ depends linearly on $(\xi_1,\xi_2)$, the conditions $B(\pm 1)=0$ remain represented by straight lines in the $(u,w)$ phase plane.

To remain consistent with the example in Fig.~\ref{fig:CircleTheCuspV1}, we again set $\xi_3=7.56$ and $\xi_4=-8.37$. When varying the amplitude, we keep the frequency fixed at $p=0.8$. Consider the point labeled $a$ in Fig.~\ref{fig:CircleTheCuspV1}, with $(\xi_1,\xi_2)=(-3.43,\,14.23)$. Using the amplitude scaling of the $\xi$-variables,
\[
(\xi_1,\xi_2,\xi_3,\xi_4)=\bigl(A\xi_{10},\,A^2\xi_{20},\,A^3\xi_{30},\,A^4\xi_{40}\bigr),
\]
we obtain a trajectory in phase space as $A$ decreases, for example from $A=1$ toward $A=0$. 
For the point labeled \(a\) with base coordinates \((\xi_{10}, \xi_{20}) = (-3.43, 14.23)\) at \(A=1\), we trace its trajectory in the \((u, w)\)-plane by evaluating the coordinate transformation along \(\xi_i(A) = A^i \xi_{i0}\) (with \(\xi_{30} = 7.56\), \(\xi_{40} = -8.37\)) at \(p = 0.8\).
with $p=0.8$ and $(\xi_{30},\xi_{40})=(7.56,\,-8.37)$.

As the amplitude or frequency varies, the lines defined by $B(\pm 1)=0$ shift, and the trajectories may therefore enter different phase regions even though the cusp region itself remains fixed. In the limit $A \to 0$, all trajectories collapse to the origin $(u,w)=(0,0)$.

\section{Conclusions}

We introduce an analytically tractable subclass of voltage-controlled generalized memristive systems that enables the systematic study of hysteresis loops exhibiting multiple self-crossing points. The model incorporates linear state relaxation governed by a characteristic time constant and a memductance non-linearly perturbed around a constant baseline admittance, with  a fourth-order polynomial voltage-dependent generating function. A parameter-independent consistency test derived from conserved quantities confirms that the formulation faithfully reproduces experimentally observed current–voltage hysteresis.
Under sinusoidal voltage excitation, a canonical mapping onto the unit circle in an auxiliary plane cleanly separates the ascending and descending branches of the loop. This geometric framework yields an explicit polynomial representation of the current in terms of nonlinear conductance and susceptance components. A novel key result is that every self-crossing point occurs precisely at a root of the susceptance polynomial, subject to a passivity constraint that confines the loop to physically admissible regions.
Exploiting the direct relation between susceptance and the derivative of conductance, we analyze the dynamics within the framework of the cusp catastrophe. Bifurcation analysis in control-parameter space identifies the cusp singularity as the critical point where the number of self-crossings changes between one and three, marking the apex of the bistability region. A  transformation to Tschirnhaus canonical coordinates then visualizes the continuous phase-space flow of the hysteresis topology as drive amplitude and frequency are varied, revealing a fixed cusp region accompanied by moving boundary constraints.
This combination of geometric mapping, polynomial susceptance analysis, and catastrophe theory provides a powerful analytical toolbox for classifying, predicting, and designing memristive devices with complex multi-lobed hysteresis characteristics.

For the sinusoidal excitation treated in this work, the pair $(V,\,dV/dt)$ adopted in the unit-circle parametrization and the pair $(V,\,\int V\,dt)$ --- in which the second coordinate is Chua's original flux variable~\cite{Chua1971} stripped of its magnetic connotation, as argued by Vongehr and Meng~\cite{Vongehr2015} --- are related by a constant of proportionality; consequently, the crossing-number classification is independent of which conjugate variable is chosen.

The topological invariants of the hysteresis loop --- the number of self-crossings and the bifurcation boundaries governing their emergence and disappearance --- are thus agnostic to whether the stimulus history is encoded through its instantaneous rate of change or through its time-integral, placing the present framework in direct contact with the flux-based language of the memristor tradition without requiring the magnetic content of the original 1971 formulation.

\bibliography{AAMultiCrossingPaperCITATIONS}

\end{document}